# Data-driven Turbulence Modeling for Separated Flows Considering Non-Local Effect

Chenyu Wu[1], Shaoguang Zhang[1], Changxin Guo[2], Yufei Zhang[1,*]

(1. School of Aerospace Engineering, Tsinghua University, 10084, Beijing, China)

(2. North China Institute of Aerospace Engineering, 065000, Langfang, China)

(*Corresponding author, zhangyufei@tsinghua.edu.cn)

**Abstract**

This study aims to enhance the generalizability of Reynolds-averaged Navier–Stokes (RANS) turbulence models, which are crucial for engineering applications. Classic RANS turbulence models often struggle to predict separated flows accurately. Recently, Data-driven machine learning approaches for turbulence modeling have been explored to address this issue. However, these models are often criticized for their limited generalizability. In this study, we address this issue by incorporating non-local effects into data-driven turbulence modeling. Specifically, we introduce a transport equation for the correction term $\beta$ of the shear stress transport (SST) model to encode non-local information along the mean streamline. The coefficients of the equation are calibrated using high-fidelity data from the NASA hump and periodic hills. The resulting model, termed the $\beta$-Transport model, demonstrates high accuracy across various separated flows outside the training set, including periodic hills with different geometries and Reynolds numbers, the curved backward-facing step, a two-dimensional bump, and a three-dimensional simplified car body. In all tested cases, the $\beta$-Transport model yields smaller or equal prediction errors compared to those from classic local data-driven turbulence models using algebraic relations for $\beta$. These results indicate improved generalizability for separated flows. Furthermore, the $\beta$-Transport model shows similar accuracy to the baseline model in basic flows including the mixing layer and the channel flow. Therefore, the non-local modeling approach presented here offers a promising pathway for developing more generalizable data-driven turbulence models.

**Keywords**: turbulence modeling, non-local effect, transport equation, data-driven approach

# 1. Introduction

Turbulence is a common occurrence in the flows crucial to engineering. However, due to its complex, chaotic, and unsteady characteristics, directly solving the Navier–Stokes equations numerically (DNS) to resolve turbulence is extremely expensive. Therefore, turbulence modeling methods are utilized to reduce computational costs. Among these methods, the Reynolds-averaged

Navier–Stokes (RANS) equations are widely used in engineering tasks due to their low computational cost compared to other turbulence modeling techniques such as large eddy simulation or DNS. The RANS equations work in conjunction with turbulence models to solve the mean flow field (pressure, velocity, and density) and turbulence quantities simultaneously. As noted by Duraisamy et al. [1], the RANS method is expected to remain a key tool in computational fluid dynamics (CFD) for engineering applications in the near future. However, studies have indicated that existing RANS turbulence models often struggle with complex separated flows common in engineering [2][3][4] scenarios, such as the stall of a high-lift device [5][6] and blunt body separation. This limitation is attributed to the numerous assumptions and empirical estimations incorporated during the construction of RANS models. Enhancing the accuracy of RANS turbulence models in complex separated flows is crucial to expanding the utility of the RANS method in engineering applications.

In recent years, data-driven methods have been successfully utilized to improve the accuracy of RANS turbulence models in separated flows [7]. Several studies have focused on enhancing the representation of Reynolds stress. For instance, Ling et al. [8] introduced a tensor-based neural network that is capable of predicting the weights of non-linear bases of Reynolds stress. Yin et al. [9] developed a unique feature engineering approach to ensure the machine-learning-predicted Reynolds stress is smoother. Additionally, Yin et al. [10] established an iterative framework that integrates the machine-learning turbulence model with CFD analysis. Another area of research has been dedicated to addressing the functional errors in turbulence model equations resulting from assumptions made by modelers. Singh et al. [11] and Parish et al. [12] utilized the field inversion and machine learning (FIML) method to quantify uncertainty in the equation of the Spalart-Allmaras model. Bidar et al. [13][14] further enhanced the FIML framework by making it open-source through the DAFoam software [15][16][17], thereby facilitating the adoption of the method. The FIML approach has gained widespread recognition over the years and has been applied to various separated flows [14][18][19][20][21][22][23], including three-dimensional ones [24].

However, there are still several limitations present in data-driven machine-learning-based turbulence models. Research by Rumsey et al. [25] highlighted that the FIML model, which relies on a neural network, did not show any improvement in separated flows that were not included in the training dataset, indicating poor generalizability. Spalart [26][27] has contended that data-driven turbulence models should maintain the same level of accuracy as traditional turbulence models in wall-attached boundary layers; however, most models fall short of achieving this standard. Moreover, complex black-box models such as neural networks and random forests can lead to significant computational costs and may lack interpretability. These drawbacks could limit the engineering applications of data-driven turbulence models. In recent studies, researchers have shifted their focus toward enhancing the generalizability of data-driven turbulence models. For instance, Tang et al. [28] attempted to identify an explicit expression for Reynolds stress rather than relying on a black-box model utilizing high-fidelity DNS data. The explicit polynomial expression for Reynolds stress exhibited good generalizability in test cases that differed from the training dataset. Additionally, Wu and Zhang [29] and He et al. [30] combined field inversion and symbolic regression methodologies to derive concise analytical expressions for the correction term of turbulence model equations. These resulting models are easily interpretable and can be applied to separated flows that are vastly different from the training scenarios. Furthermore, efforts have been made to incorporate constraints from traditional turbulence models, particularly those relevant to attached boundary layers, into the machine-learning phase. By adhering to these constraints, data-driven models have been developed to ensure the accuracy of flows attached to the wall, as demonstrated by Bin et al. [31][32]. Wu et al. [33] introduced a conditioned field inversion approach to preserve the calibration of traditional turbulence models in attached boundary layers, thereby improving the robustness of data-driven turbulence models in simple attached flows.

In all of the aforementioned research, the machine-learning correction term is local, meaning that the value of the correction term at point **x** is determined by flow features such as strain rate at point **x**. However, as pointed out by Spalart [27], one of the key characteristics of turbulence is its non-local nature. This implies that the turbulence state at point **x** is not only influenced by flow features at **x** but also by the characteristics of the flow at other points in the surrounding region. Therefore, a non-local machine-learning correction term could potentially be more representative and may capture more information regarding flow separation, thereby improving the generalizability of the data-driven turbulence model. The difference between a local correction term and a non-local correction term is illustrated in Figure 1. Zhou et al. [34] and Han et al. [35] have pioneered advancements in non-local turbulence modeling by introducing the vector-cloud neural network to incorporate non-local information. However, for engineering applications, a more lightweight and interpretable non-local model is preferred over a black-box neural network. Several studies [36][37][38] have derived transport equations for the eddy viscosity $v_T$ or its correction term to account for the non-local effect along the streamlines. Although a transport equation is indeed more concise than a deep neural network, the transport equations derived in these studies are either purely from empirical observations or from existing Reynolds stress models, without utilizing the extensive high-fidelity data on separated flows accumulated over the years [36][37][38]. Moreover, these transport equations are not specifically tailored for massively separated flows, which are the focus of our research interest.

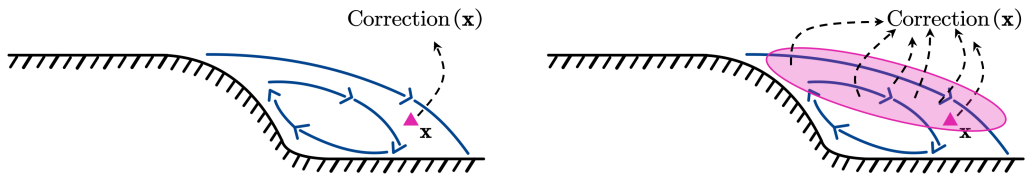

Figure 1. The difference between local correction (left) and non-local correction (right). The purple triangle is the point **x** under consideration. In the right panel, the purple oval represents the adjacent area of the point **x** that the non-local correction term might depend on.

In this paper, we aim to enhance the performance of data-driven turbulence models in separated flows by incorporating non-local characteristics. We selected Menter's Shear-Stress-Transport (SST) model [39] as the model for correction. A correction term $\beta$ was introduced to the SST model's equations, and a transport equation for $\beta$ was developed based on our experience to account for non-local effects along the streamline. The parameters of the transport equation were calibrated using simple data-driven techniques with evolutionary algorithms against the training set. We avoided using black-box models to ensure interpretability, portability, and computational efficiency. The resulting model, named the $\beta$-Transport model, demonstrated superior or equal accuracy in a series of separated flows different from the training set compared to the local data-driven turbulence model presented in our previous study [33]. Therefore, incorporating non-local effects into data-driven turbulence models could potentially enhance the generalizability of the model in separated flows.

## 2. Formulation of the non-local model

In this section, we introduce the developed non-local model. First, we provide a brief description of the baseline traditional turbulence model that is being corrected, namely the SST-2003 model. Next, we formulate the transport equation for the correction term $\beta$ based on our experience and the explicit local models of $\beta$ found in the literature. We then proceed to describe the physical meaning of each term in the equation for $\beta$.

Given the extensive use of the SST 2003 model (abbreviated as SST) in engineering applications, we have selected it as the baseline model for correction. The SST model includes two transport equations for turbulent kinetic energy $k$ and specific dissipation rate $\omega$ for incompressible flow with constant density. The equations are:

$$\frac{\partial k}{\partial t} + u_j \frac{\partial k}{\partial x_j} = P_k - \beta^* k \omega + \frac{\partial}{\partial x_j}\left[(\nu + \sigma_k \nu_T)\frac{\partial k}{\partial x_j}\right]$$

$$\frac{\partial \omega}{\partial t} + u_j \frac{\partial \omega}{\partial x_j} = \frac{\gamma}{\nu_T} P_k - \theta \omega^2 + \frac{\partial}{\partial x_j}\left[(\nu + \sigma_\omega \nu_T)\frac{\partial \omega}{\partial x_j}\right] + 2(1 - F_1)\frac{\sigma_{\omega 2}}{\omega}\frac{\partial k}{\partial x_j}\frac{\partial \omega}{\partial x_j} \quad (1)$$

The production of turbulent kinetic energy, denoted as $P_k$ is defined as:

$$P_k = \min\left(\tau_{ij}\frac{\partial u_i}{\partial x_j}, 10\beta^* k \omega\right)$$

$$\tau_{ij} = 2\nu_T S_{ij} - \frac{2}{3}k\delta_{ij} \quad (2)$$

$$S_{ij} = \frac{1}{2}(u_{j,i} + u_{i,j})$$

In the context of this research, $\tau_{ij}$ represents the Reynolds stress, $\nu_T$ stands for the eddy viscosity, and $S_{ij}$ denotes the strain rate. $u_{i,j}$ corresponds to the partial derivative of the velocity component along the $i^{th}$ direction with respect to the $x_j$ direction. Furthermore, the eddy viscosity $\nu_T$ is determined as follows:

$$\nu_T = \frac{a_1 k}{\max(a_1 \omega, SF_2)} \quad (3)$$

$S$ is the magnitude of the strain rate: $S = \sqrt{2S_{ij}S_{ij}}$. The blending functions $F_1$ and $F_2$ are defined as follows:

$$F_1 = \tanh(\xi_1^4), \xi_1 = \min\left[\max\left(\frac{\sqrt{k}}{\beta^*\omega d}, \frac{500\nu}{d^2\omega}\right), \frac{4\sigma_{\omega 2}k}{CD_{k\omega}d^2}\right]$$

$$CD_{k\omega} = \max\left(2\sigma_{\omega 2}\frac{1}{\omega}\frac{\partial k}{\partial x_j}\frac{\partial \omega}{\partial x_j}, 10^{-10}\right)$$

$$F_2 = \tanh(\xi_2^2)$$

$$\xi_2 = \max\left(\frac{2\sqrt{k}}{\beta^*\omega d}, \frac{500\nu}{d^2\omega}\right)$$

(4)

The distance $d$ at point $\mathbf{x}$ is defined as the distance from $\mathbf{x}$ to the nearest wall. In Eq. (1) to (4), $\beta^*$, $a_1$, and $\sigma_{\omega 2}$ are constants. Other parameters are specified as follows:

$$\gamma = F_1\gamma_1 + (1-F_1)\gamma_2, \theta = F_1\theta_1 + (1-F_1)\theta_2$$

$$\sigma_k = F_1\sigma_{k1} + (1-F_1)\sigma_{k2}, \sigma_\omega = F_1\sigma_{\omega 1} + (1-F_1)\sigma_{\omega 2}$$

(5)

$\gamma_1, \gamma_2, \sigma_{k1}, \sigma_{k2}, \sigma_{\omega 1}$, and $\sigma_{\omega 2}$ are all constants. The specific values can be referenced in [39].

Although the SST model is widely used, it often proves unreliable in handling complex separated flows [2][5][6][40][41]. Hellsten [42] has identified a potential source of error in the scale-determining equations (the transport equations of $k$ and $\omega$) that can lead to such failures. Recent studies [41][42] have indicated that RANS models tend to underestimate the eddy viscosity $\nu_T$ in the separated shear layer between the retarded flow in the separation bubble and the mainstream. This is primarily due to the presence of a real massive separation that exhibits a slightly unsteady oscillating shear layer, resulting in strong momentum diffusion that can only be accurately predicted with a large $\nu_T$.

Several researchers [29][33][41] have proposed correcting the destruction term of the $\omega$ equation to account for significant separations, as the main source of error arises from the scale-determining equations.

$$\frac{\partial \omega}{\partial t} + u_j\frac{\partial \omega}{\partial x_j} = \frac{\gamma}{\nu_T}P_k - B\theta\omega^2 + \frac{\partial}{\partial x_j}\left[(\nu + \sigma_\omega\nu_T)\frac{\partial \omega}{\partial x_j}\right] + 2(1-F_1)\frac{\sigma_{\omega 2}}{\omega}\frac{\partial k}{\partial x_j}\frac{\partial \omega}{\partial x_j} \quad (6)$$

The correction term $B$ is a spatially varying non-dimensional field that is correlated with mean flow quantities such as $S$ and $P_k$. It can take the form of an analytical expression or a machine learning model. $B$ can be derived from experience [41] or through purely data-driven machine learning methods [29][33]. Various typical forms of $B$ are presented in Table 1. In this study, we specifically focus on machine-learning models with analytical forms or correction terms derived empirically, as these provide good interpretability to assist in our modeling process.

Table 1. Some typical forms of $B$

| Definition of $B$ | The physical meaning of the correction |
|---|---|
| $B = f_d \left\{ \min \left( 12, \max \left[ 1, \left( 4\frac{P_k}{\varepsilon} - 5 \right) \right] \right) \right\}$ $+ (1 - f_d)$ $\varepsilon = \beta^* k\omega, f_d = 1 - \tanh([8r_d]^3)$ $r_d = \frac{\nu + \nu_T}{\kappa^2 d^2 \sqrt{u_{i,j} u_{i,j}}}$ | The correction term, as derived empirically from [41]. $f_d$ equals 0 in the attached boundary layer and equals 1 elsewhere. When incorporated in the SST model, this correction is called the SST-sf model. Consequently, $f_d$ can deactivate the correction term in the attached boundary layer to preserve the accuracy of the baseline SST model. $P_k/\varepsilon$ is approximately proportional to $S_{ij}S_{ij}/\omega^2$, which increases as the shear increases. It helps to distinguish the region with strong shear. The definition of $B$ states that it deviates from 1 when $P_k/\varepsilon > 1.5$ outside the boundary layer and remains 1 inside the boundary layer. |
| $B = \left( \frac{P_k}{\varepsilon} - 0.244 \right) \tanh \left( \frac{\lambda_2 \lambda_5}{Re_\Omega} \right) + 1$ $\lambda_2 = tr(\Omega_{ik}\Omega_{kj}), \lambda_5 = tr(\Omega_{ik}\Omega_{kl}S_{lm}S_{mj})$, $Re_\Omega = \frac{\sqrt{\Omega_{ij}\Omega_{ij}}d^2}{\nu}, \Omega_{ij} = \frac{1}{2}(u_{j,i} - u_{i,j})$ | This correction term was developed in [29] using the field inversion and machine learning method [12]. It is an analytical expression derived by the symbolic regression algorithm [43] from available data. The corrected model is called the SST-SR model. Similar to [41], $P_k/\varepsilon$ is used as an indicator of strong shear. $\tanh(\eta)$ can be viewed as a 'switch' of the correction term. It activates the correction in the separated shear flow and deactivates it in the bottom of the boundary layer. |
| $B = \beta f_d + 1$ $\beta = \min(0.00435\lambda_2^2, 3.8)$ | This term is also a machine-learning model trained by symbolic regression [33]. The model is called the SST-CND model. Different from [29], it uses the conditioned field inversion method to ensure the correction term does not negatively impact the accuracy of the attached boundary layer. Similar to [41], the shielding function $f_d$ is used here. |

Table 1 indicates that $B$ is frequently chosen to be positively correlated with the shear of the mean flow ($P_k/\varepsilon, \lambda_2$, etc.). This adjustment serves to enhance the destruction of $\omega$ in the separated shear layer between the retarded flow and the mainstream, leading to a decrease in $\omega$. In accordance with Eq. (3), the eddy viscosity $\nu_T$ increases as $\omega$ decreases in the separated shear layer ($F_2$ is zero outside the boundary layer [44]). An increased $\nu_T$ indicates a heightened momentum diffusion in the separated shear layer, thereby correcting the predictions of the SST model in cases of extensive separation. This is illustrated in Figure 2, where the baseline SST model underestimates the $\nu_T$ within the separated shear layer (denoted by the region enclosed by a red dashed line), leading to an inaccurately large separation. In contrast, the SST-CND model [33] in Table 1 exhibits a larger $\nu_T$, resulting in a more precise prediction of the separation size. Additionally, empirically constructed shielding functions ($f_d$) or those generated using machine-learning methods ($\tanh(\lambda_2\lambda_5/Re_\Omega)$) are commonly employed to deactivate the correction term in the attached boundary layer. This is because traditional turbulence models are proficient in handling simple attached flows, and additional machine-learning corrections may compromise their accuracy [25].

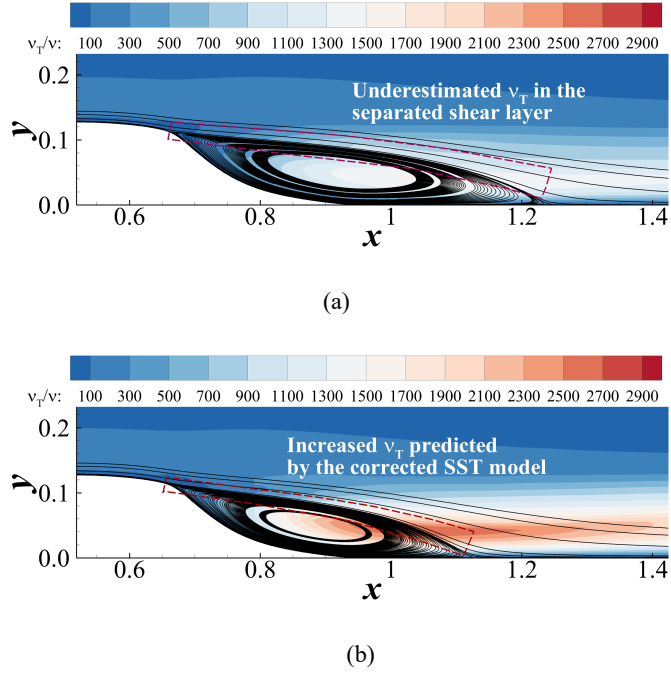

Figure 2. The $\nu_T/\nu$ predicted by (a) the SST model and (b) SST-CND model [33].

All the expressions and machine learning models in Table 1 only use local flow variables as input. In other words, $B$ at a point **x** is determined by the flow variables at **x**. However, Spalart [27] suggests that turbulence is essentially non-local. Therefore, it is reasonable to hypothesize that the correction term $B$, which is associated with the flow turbulence in the separated shear layer, should also be non-local to more accurately describe the turbulence. Building on the pioneering works in Refs. [36][37][38], we propose a transport equation to account for the non-locality of the correction term $B$. The proposed transport equation is integrated with the baseline SST model, resulting in the development of the $\beta$-Transport model.

$$\frac{\partial k}{\partial t} + u_j \frac{\partial k}{\partial x_j} = P_k - \beta^* k\omega + \frac{\partial}{\partial x_j}\left[(\nu + \sigma_k \nu_T)\frac{\partial k}{\partial x_j}\right]$$

$$\frac{\partial \omega}{\partial t} + u_j \frac{\partial \omega}{\partial x_j} = \frac{\gamma}{\nu_T} P_k - B\theta\omega^2 + \frac{\partial}{\partial x_j}\left[(\nu + \sigma_\omega \nu_T)\frac{\partial \omega}{\partial x_j}\right] + 2(1-F_1)\frac{\sigma_{\omega 2}}{\omega}\frac{\partial k}{\partial x_j}\frac{\partial \omega}{\partial x_j} \quad (7)$$

$$\frac{\partial \beta}{\partial t} + u_j \frac{\partial \beta}{\partial x_j} = C_{\beta 1} \frac{\nu_T}{d^2} \frac{\sqrt{\Omega_{ij}\Omega_{ij}}}{\beta^* \omega} f_d \left(1 - \frac{1}{4}\beta\right) - C_{\beta 2} \omega \beta^m + \frac{\partial}{\partial x_j}\left[(\nu + \sigma_\beta \nu_T)\frac{\partial \beta}{\partial x_j}\right] \quad (8)$$

$$B = C_{\beta 3} \beta f_d + 1 \quad (9)$$

It should be noted that Eq. (8) is formulated based purely on empirical and mathematical arguments, which will be elaborated upon in the subsequent paragraphs. The precise physical interpretation of each term in Eq. (8) may be unclear as the 'ideal' distribution and the physical significance of $\beta$ are not yet clearly defined. In existing literature, $\beta$ is simply regarded as a numerical representation of the model's uncertainty [11][12]. However, the primary objective of this paper is to demonstrate that a data-driven turbulence model can be developed by incorporating transport equations for the correction term to account for non-local effects. Although the theoretical and physical significance of this transported correction term is a pertinent topic, it is beyond the scope of the present study.

The non-locality is represented by the non-dimensional parameter $\beta$, which is encoded as $B$ in Eq. (8) and (9). Constants $C_{\beta 1}, C_{\beta 2}, C_{\beta 3}, \sigma_\beta, m$ are involved in this representation. Building upon the work of [33][41], a shielding function $f_d$ (as detailed in Table 1) is introduced in Eq. (9) to ensure that the correction term $B$ does not significantly affect the boundary layer (where $B = 1$, as $f_d$ is zero in the boundary layer). The first term on the right side of Eq. (8) represents the production term of $\beta$.

$$P_\beta = C_{\beta 1} \frac{\nu_T}{d^2} \frac{\sqrt{\Omega_{ij}\Omega_{ij}}}{\beta^* \omega} f_d (1 - \frac{1}{4}\beta) \qquad (10)$$

$\nu_T/d^2$ has the dimension of inverse time. It approaches zero in free shear flows as $d$ tends to infinity when there is no solid wall in the domain. In other words, the production term is switched off in free shear flows, maintaining the baseline SST model's behavior in these flows. The ratio of $\sqrt{\Omega_{ij}\Omega_{ij}}/(\beta^* \omega)$ represents the non-dimensional rotation rate. In thin shear flows like separated shear layers, the rotation rate is approximately equal to the strain rate, making it a good indicator of flow shear magnitude. We do not use the strain rate directly to measure shear because in flows with intense compression or expansion (such as flow around the stagnation point of an airfoil), the strain rate is very large with no correction needed, while the rotation rate approaches zero. $\sqrt{\Omega_{ij}\Omega_{ij}}/(\beta^* \omega)$ in Eq. (10) signifies the production of the correction term in situations of intense shear flow, aligning with previous corrections for massive separations as shown in Table 1. The function $f_d$ acts as a shield, turning off the production of $\beta$ in the attached boundary layer to maintain the baseline model's behavior. The $(1 - \beta/4)$ term in Eq. (10) deactivates the production of $\beta$ when $\beta > 4$, thus limiting $\beta$ to be approximately less than 4. The second term on the right-hand side of Eq. (8) denotes the destruction term of $\beta$.

$$D_\beta = C_{\beta 2} \omega \beta^m \qquad (11)$$

Where $C_{\beta 2}$ and $m$ are constants that are both larger than zero. Eq. (11) ensures that $\beta$ approaches zero in the mainstream where the turbulence intensity is low ($\omega$ often takes a relatively large value in this region). Additionally, the last term in Eq. (8) represents a diffusion term that accounts for the diffusion of the correction term caused by eddy viscosity. In summary, the specific form of Eq. (8) presented here is designed to minimize the influence of $\beta$ on simple flows, such as free shear flows, channel flows, and flows that are devoid of turbulence.

The fundamental concept of the $\beta$-Transport model is comparable to the models listed in Table 1. It aims to increase the correction term in the separated shear layer to enhance the destruction of $\omega$, ultimately leading to an increase in the eddy viscosity $\nu_T$. The primary distinction between the newly proposed non-local transport equation formulation and previous models lies in the correction term $B$ at a given point $\mathbf{x}$, which can exceed 1 when there is strong shear upstream or beside $\mathbf{x}$, even if the shear itself (i.e., $\sqrt{\Omega_{ij}\Omega_{ij}}/(\beta^*\omega)$) is weak at $\mathbf{x}$. This is because Eq. (8) permits the enhanced $\beta$ upstream of $\mathbf{x}$ caused by strong shear to be conveyed to $\mathbf{x}$ by the mean flow, regardless of the shear at $\mathbf{x}$. Furthermore, it enables the increased $\beta$ to be diffused from the surrounding region of $\mathbf{x}$. This significant distinction is depicted in Figure 3, showcasing the inherent non-locality of the $\beta$-Transport model introduced by Eq. (8).

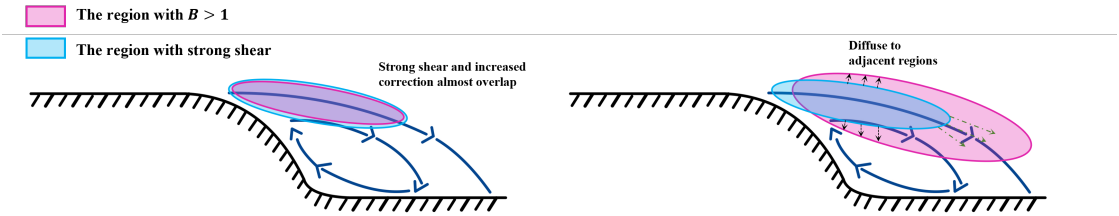

Figure 3. The key difference between the local model of the correction term and the non-local transport equation-based model.

The functional forms of Eq. (8) and (9) have been developed empirically. Alternatively, it is feasible to derive them from a data-centered machine-learning framework, such as the FIML framework [21]. However, this is outside the scope of the current paper, as our main objective is to demonstrate how incorporating non-local characteristics can improve the model's performance in unseen separated flows. In the subsequent section, we will employ a simple data-driven approach to train the model, specifically by calibrating the model constants in Eq. (8) and (9).

# 3. Data-driven model calibration

Following the suggestion of Duraisamy et al. [1], we consider the data-driven method as an approach to calibrate the model. Specifically, we utilize an evolutionary algorithm [45] to optimize the model constants in Eq. (8) and (9). The high-fidelity data of the periodic hill [46] and the NASA hump [47] are utilized as the training data during the optimization process. The model constants are adjusted in order to minimize the difference in the flow field between the prediction of the $\beta$-Transport model and the high-fidelity data. In the subsequent subsections, we will provide a detailed explanation of the data-driven calibration process.

## 3.1. Optimization method

In this paper, we utilize the differential evolution (DE) method, as proposed in [48], as the fundamental optimization algorithm. The DE method is known for its efficiency and effectiveness in locating the global optimum without the need for gradient information of the objective function. It operates by iteratively updating a population of $K$ individuals for $N$ generations using heuristic techniques such as mutation and crossover. Let $\boldsymbol{w}_i^G$ ($i = 1, \cdots, K; G = 0, \cdots, N$) represent the $i^{th}$ individual of the $G^{th}$ generation. $\boldsymbol{w}_i^G$ is a $D$-dimensional vector whose components correspond to the parameters to be optimized, such as the model constants $C_{\beta 1}, C_{\beta 2}, C_{\beta 3}, \sigma_\beta, m$ in Eq. (8) and (9). The objective function value associated with vector $\boldsymbol{w}_i^G$ is denoted by $J(\boldsymbol{w}_i^G)$. For demonstration purposes, we focus on a single objective function in this study, although the optimization algorithm is capable of handling multiple objectives. The DE method updates the $G^{th}$ population to generate the $(G + 1)^{th}$ population using the following approach:

- **Mutation**: For each individual $\boldsymbol{w}_i^G$, calculate a mutant vector $\boldsymbol{v}_i^G$.

$$\boldsymbol{v}_i^G = \boldsymbol{w}_{r_1}^G + F(\boldsymbol{w}_{r_2}^G - \boldsymbol{w}_{r_3}^G) \tag{12}$$

Where $r_1, r_2$, and $r_3$ are three different indices randomly selected from $1, \cdots, K$. $F$ is a positive constant.

- **Crossover**: For each individual $\boldsymbol{w}_i^G$ and its mutant vector $\boldsymbol{v}_i^G$, calculate a crossover vector $\boldsymbol{u}_i^G$. Let the $j^{th}$ component of $\boldsymbol{u}_i^G$ be $u_{ij}^G$. $u_{ij}^G$ is calculated as:

$$u_{ij}^G = \begin{cases} v_{ij}^G, \text{if } (\text{rand}(j) < CR) \text{ or } j = \text{randint}(i) \\ w_{ij}^G, \text{if } (\text{rand}(j) \geq CR) \text{ and } j \neq \text{randint}(i) \end{cases} \tag{13}$$

Here, $CR \in [0,1]$ is the crossover rate specified by the user, $\text{rand}(j)$ is a random number between 0 and 1, and $\text{randint}(i)$ is a random integer chosen from $1, \cdots, D$. In summary, Eq. (13) randomly selects the component of $\boldsymbol{u}_i^G$ from the components of $\boldsymbol{v}_i^G$ and $\boldsymbol{w}_i^G$, ensuring that at least one component of $\boldsymbol{u}_i^G$ is from $\boldsymbol{v}_i^G$.

- **Selection**: Compute $J(\boldsymbol{u}_i^G)$ and $J(\boldsymbol{w}_i^G)$, if $J(\boldsymbol{u}_i^G) < J(\boldsymbol{w}_i^G)$, then set $\boldsymbol{w}_i^{G+1} = \boldsymbol{u}_i^G$. Otherwise, set $\boldsymbol{w}_i^{G+1} = \boldsymbol{w}_i^G$.

The in-house DE optimization program SAMO is utilized in this study. In addition to the basic DE method described earlier, SAMO incorporates the use of a radial-basis function (RBF) response surface to enhance the elite individuals in the optimization process and improve the diversity of the population [45]. The RBF response surface is a straightforward machine-learning model that correlates the design variable vector $\boldsymbol{w}_i^G$ with the value of objective function $J(\boldsymbol{w}_i^G)$. The objective value estimated by the RBF response surface is denoted as $\hat{J}(\boldsymbol{w}_i^G)$. It is important to note that $\hat{J}(\boldsymbol{w}_i^G)$ serves as an approximation of $J(\boldsymbol{w}_i^G)$. The response surface is continuously updated during the DE optimization process as new individuals are evaluated. During each optimization iteration, a 'local' optimization is carried out utilizing the approximate objective $\hat{J}$ predicted by the response surface. A set of individuals ($\boldsymbol{c}_i^G$) with optimized $\hat{J}$ is then generated. The new generation of the population $\boldsymbol{w}_i^{G+1}$ is selected from $\boldsymbol{c}_i^G$ and $\boldsymbol{u}_i^G$ (generated by the basic DE process). The simple machine-learning response surface can rapidly estimate the objective compared to traditional CFD, thereby enhancing the efficiency of the search by incorporating a more diverse set of elite individuals $\boldsymbol{c}_i^G$ for selection. The overall flowchart of the optimization process is illustrated in Figure 4. For further details on the selection criteria, readers are encouraged to refer to [45].

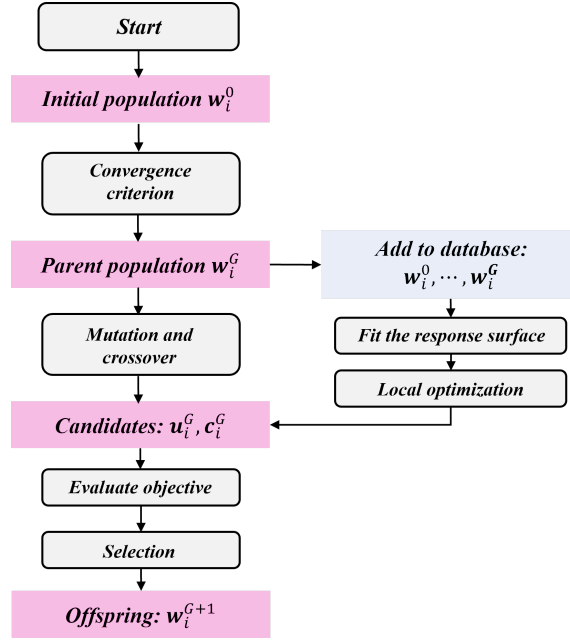

Figure 4. The flowchart of the optimization process.

## 3.2. Optimization setup

### 3.2.1. Optimization parameters

In the data-driven calibration process, the vector of design variables is defined as:

$$\boldsymbol{w} = (C_{\beta 1}, C_{\beta 2}, C_{\beta 3}, \sigma_\beta, m) \tag{14}$$

The design variable boundaries are outlined in Table 2. In order to conduct multi-objective optimization, two objectives and one constraint have been defined.

$$\min_{\boldsymbol{w}} J_1(\boldsymbol{w}) = 20|x_{\text{rth,PHLL}} - x_{\text{rth,PHLL}}^{\text{LES}}| + 2\sum_{j=1}^{N}|C_f^k(\boldsymbol{w}) - C_{f,\text{LES}}^k|$$

$$J_2(\boldsymbol{w}) = 50|x_{\text{rth,HUMP}} - x_{\text{rth,HUMP}}^{\text{Exp}}| + 2\sum_{k=1}^{N}|C_f^k(\boldsymbol{w}) - C_{f,\text{Exp}}^k| \tag{15}$$

$$s.t. MAE_{zpg} = \frac{1}{N}\sum_{j=1}^{N}\frac{|C_f^j(\boldsymbol{w}) - C_{f,SST}^j|}{C_{f,SST}^j} \times 100\% < 3.0\%$$

The reattachment points $x_{\text{rth,PHLL}}$ and $x_{\text{rth,HUMP}}$ refer to the locations where flow separation reattaches on the periodic hill (Re = 10595) and the NASA hump, respectively. $x_{\text{rth,PHLL}}^{\text{LES}}$ and $x_{\text{rth,HUMP}}^{\text{Exp}}$ are the reattachment points (ground truth) obtained by LES through Large Eddy

Simulation (LES) [49] and experimental data [50]. The second term of $J_1$ and $J_2$ measures the difference between the friction coefficient ($C_f$) given by the $\beta$-Transport model and the high-fidelity data. $C_f^k(\mathbf{w})$ represents the interpolated value of the $\beta$-Transport model's prediction at the $k^{th}$ location where high-fidelity data is available. In Eq.(15), $MAE_{zpg}$ is the mean absolute error of $C_f$ between the SST model's prediction and the $\beta$-Transport model in the zero pressure-gradient turbulent flat plate case. $N$ represents the number of faces along the plate surface. $C_f^j(\mathbf{w})$ and $C_{f,SST}^j$ represent the friction coefficient in the $j^{th}$ face of the flat plate predicted by the $\beta$-Transport model and the SST model, respectively. The objective of Eq. (15) is to minimize the difference in the reattachment point predicted by the $\beta$-Transport model and the high-fidelity data, while also ensuring the mean absolute error in the zero pressure-gradient attached boundary layer remains below 3.0% to maintain the baseline accuracy of the SST model in basic attached flows. The multipliers (20 and 50) in Eq. (15) are used to scale the values of $J_1$ and $J_2$. Figure 5 displays the streamline plots and skin friction coefficients $C_f$ of the periodic hill and the NASA hump, obtained from high-fidelity data and used as the ground truth. The reattachment point is defined as the location where the $C_f$ curve intersects $C_f = 0$ and $dC_f/dx > 0$.

The optimization process is scheduled to run for 25 generations, with a population size of 12 individuals. The initial population is generated using the Latin-hyper-cube sampling method with the Python package PyDOE [51].

Table 2. Lower and upper bounds of the design variables

|  | Lower bound | Upper bound |
| --- | --- | --- |
| $C_{\beta 1}$ | 12.0 | 80.0 |
| $C_{\beta 2}$ | 2.0 | 12.0 |
| $C_{\beta 3}$ | 0.7 | 4.5 |
| $\sigma_\beta$ | 0.25 | 2.0 |
| $m$ | 0.3 | 0.6 |

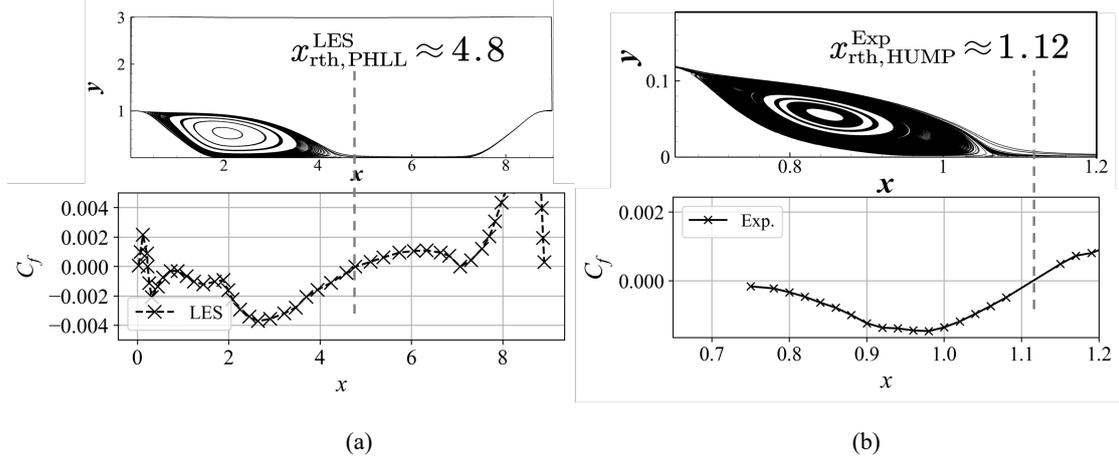

(a)                                              (b)

Figure 5. The flow separation and the skin friction coefficient plot of (a) the periodic hill and (b) the NASA hump.

### 3.2.2. Numerical settings for objective function evaluation

In this paper, we utilized OpenFOAM's SimpleFoam solver [52] to simulate the flow and calculate the value of the objective function function. The solver is a pressure-based incompressible solver. We employed second-order numerical schemes for unstructured mesh to compute the face flux. The grids used for the periodic hill (for objective $J_1$, using $180 \times 80$ cells), the NASA hump (for objective $J_2$, using $576 \times 152$ cells), and the zero-pressure gradient flat plate (for the constraint, using $136 \times 96$ cells) are shown in Figure 6. In all the meshes, the value of $\Delta y^+$ is maintained to be smaller than 1. Specifically, for the turbulent flat plate, $\Delta y^+$ is approximately 0.2 along the wall.

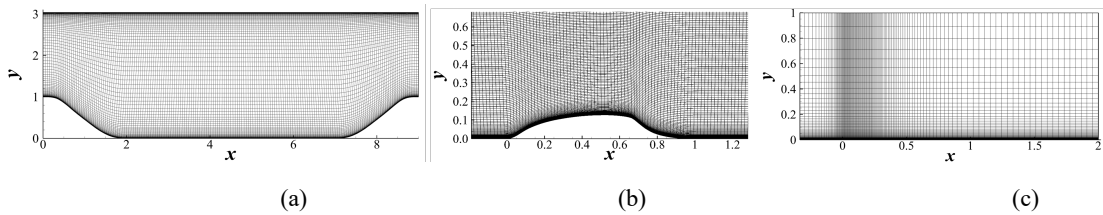

(a)                              (b)                               (c)

Figure 6. The computational grids for (a) the periodic hill, (b) the NASA hump, and (c) the zero pressure gradient flat plate.

## 3.3. Optimization results

All the individuals produced in the optimization process are plotted in the $J_1 - J_2$ space in Figure 7. The individuals on the Pareto front are marked in dark red, and the initial population is indicated by blue triangles. The individuals on the Pareto front are not 'dominated' by any other individual. In other words, for each $x$ on the Pareto front, there does not exist an individual whose two objectives are both smaller than $x$. Therefore, all the individuals on the Pareto front can be considered as 'optimized'. Comparing with the initial population, the Pareto front shows significant movement towards (0,0), indicating the effectiveness of our optimization.

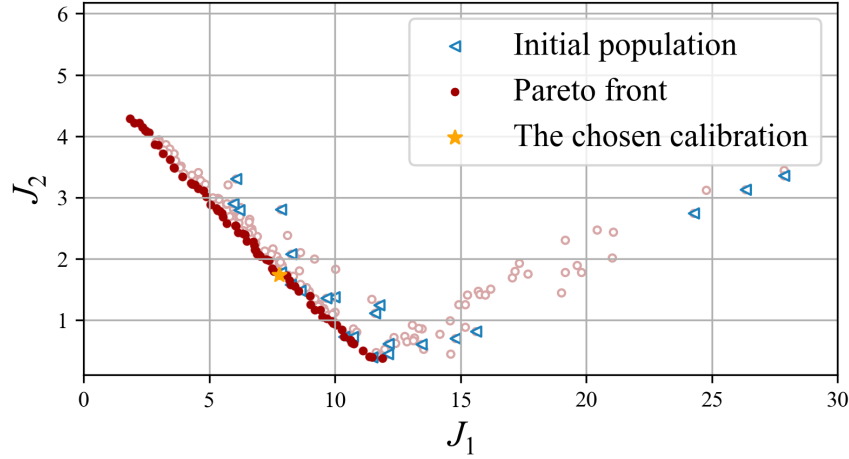

Figure 7. The individuals produced in the optimization process.

The final calibration of the parameters is determined as follows. The distance is calculated for all individuals in the Pareto front.

$$d_p = \sqrt{\left(\frac{J_1 - J_{1,\min}}{J_{1,\max} - J_{1,\min}}\right)^2 + \left(\frac{J_2 - J_{2,\min}}{J_{2,\max} - J_{2,\min}}\right)^2} \quad (16)$$

The minimum and maximum of $J_1$ and $J_2$ are determined across the entire Pareto front. The individual within the Pareto front with the smallest $d_p$ is considered the optimal balance between $J_1$ and $J_2$, assuming equal importance of both objectives. This individual is selected as the final calibration. The errors in the reattachment point and $C_f$ are detailed in Eq. (17), and the calibrated model parameters are presented in Table 3.

$$\left|x_{\text{rth,PHLL}} - x_{\text{rth,PHLL}}^{\text{LES}}\right| = 0.36, \left|x_{\text{rth,HUMP}} - x_{\text{rth,HUMP}}^{\text{Exp}}\right| = 0.023,$$
$$MAE_{zpg} = 2.77\% \quad (17)$$

Table 3. The value of the calibrated parameters

| $C_{\beta 1}$ | $C_{\beta 2}$ | $C_{\beta 3}$ | $\sigma_\beta$ | $m$ |
|---|---|---|---|---|
| 76.06 | 2.00 | 2.72 | 0.34 | 0.52 |

## 3.4. Application of the calibrated model on the training set

In this section, we utilize the calibrated $\beta$-Transport model to analyze the cases incorporated in the objective function and the constraint in the optimization process. We also make a comparison between the current non-local data-driven $\beta$-Transport model and the local data-driven SST-CND model [33], as outlined in Table 1. It is important to note that throughout the discussions below, $\beta$ is plotted instead of the correction term $B$ for better clarity ($\beta = 0$ indicates no correction, while $\beta > 0$ indicates correction). Grid convergence studies are conducted for both the training cases.

### 3.4.1. Periodic hill, $Re = 10595$

Three levels of meshes are applied in this training case. The coarse mesh has $180 \times 80$ cells, which is identical to the mesh used in the calibration process. The medium and the fine mesh have $256 \times 112$ cells and $360 \times 160$ cells, respectively. Figure 8 shows the separation bubble and the distribution of the correction term for the $\beta$-Transport model and the SST-CND model obtained on the medium mesh. In terms of the size and shape of the separation bubble, the $\beta$-Transport model aligns well with the LES data, while both the SST and the SST-CND models overpredict the separation size. The SST-CND model and the $\beta$-Transport model both increase the correction term $\beta$ in the separated shear layer. Due to the non-local transportation effect inherent in the $\beta$-Transport model, the correction generated by it extends further downstream and diffuses wider perpendicular to the flow direction. The velocity profiles in Figure 9 further confirm that the $\beta$-Transport model best matches the LES data. The profiles of the $\beta$-Transport model given by different meshes almost overlap, indicating good grid convergence. While the SST-CND model also predicts the separation more accurately compared to the baseline SST model, it overestimates the velocity defect when compared to the $\beta$-Transport model. $C_f$ distribution in Figure 10 suggests that the $\beta$-Transport model captures the reattachment point and the peak $C_f$ with better accuracy.

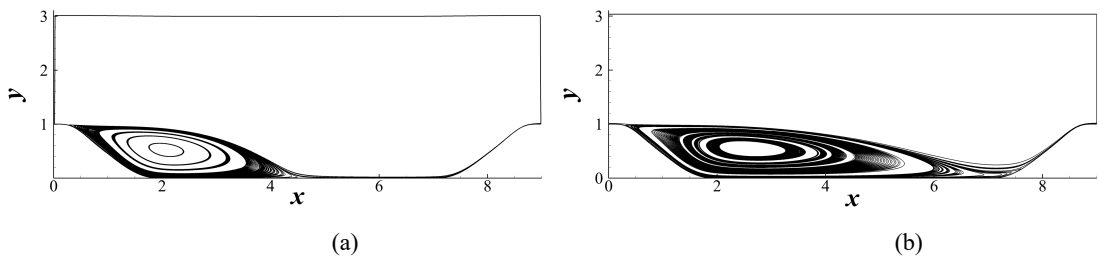

(a)                        (b)

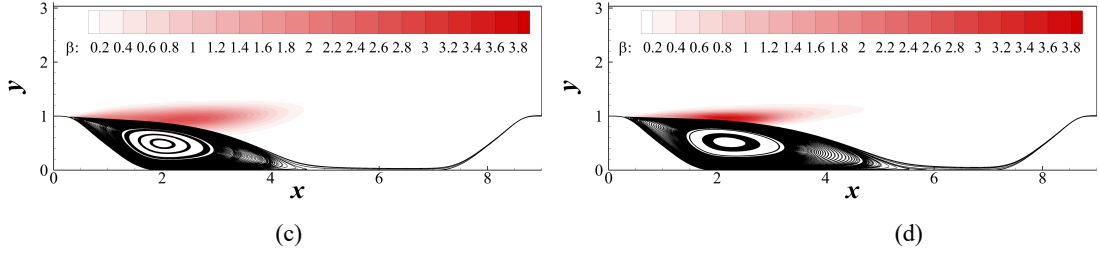

Figure 8. The separation bubble predicted by (a) LES [46], (b) the SST model, (c) the $\beta$-Transport model, (d) and the SST-CND model; for the results of the $\beta$-Transport model and the SST-CND model, the correction term $\beta$ is also plotted.

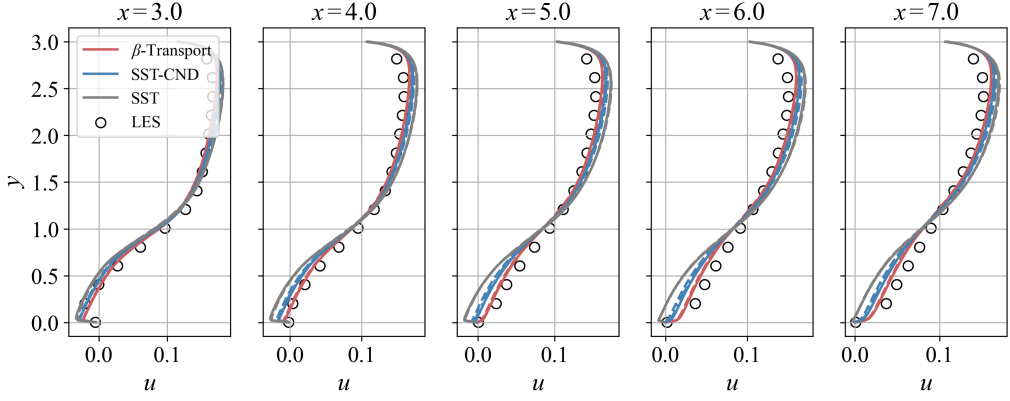

Figure 9. The velocity profiles at the different locations. The dashed lines, dash-dot lines, and solid lines represent the results of the coarse mesh, medium mesh, and fine mesh, respectively.

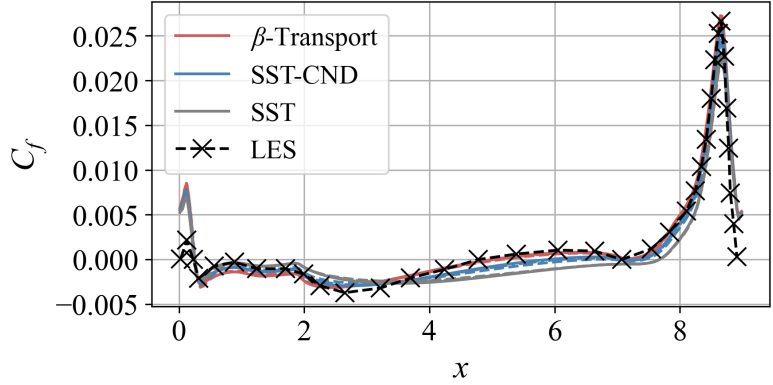

Figure 10. The skin friction coefficient distribution predicted by different models. The dashed lines, dash-dot lines, and solid lines represent the result of the coarse mesh, the medium mesh, and the fine mesh, respectively.

Let the mean squared error (MSE) of the velocity be defined as:

$$MSE = \frac{1}{M}\sum_{k=1}^{M}|\boldsymbol{u}_k^{\text{RANS}} - \boldsymbol{u}_k^{\text{True}}|^2 \qquad (18)$$

$\boldsymbol{u}_k^{\text{RANS}}$ represents the velocity predicted by the RANS method in the $k^{th}$ cell, while $\boldsymbol{u}_k^{\text{True}}$ represents the ground truth value of the velocity (obtained from LES) interpolated to the $k^{th}$ cell.

Table 4 presents the MSE of the two data-driven models in comparison to the MSE of the SST model. The results demonstrate that the $\beta$-Transport model achieves a 50% reduction in MSE when compared to the SST-CND model, indicating its superior capability in predicting separated flow in this particular case.

Table 4. The velocity MSE / the velocity MSE of the SST model

|  | SST | SST-CND | $\beta$-Transport |
|---|---|---|---|
| MSE/MSE$_{SST}$ | 100% | 31.50% | **13.29%** |

### 3.4.2. NASA hump

Three sets of grids are utilized for this case study. The coarse grid is the same as the grid used during the training process, which consists of $152 \times 576$ cells. The medium grid and the fine grid contain $216 \times 816$ cells and $304 \times 1156$ cells, respectively. The streamline plots depicted in Figure 11 indicate that both the $\beta$-Transport and SST-CND models accurately predict the size of the separation bubble, closely matching the high-fidelity LES simulation [47], whereas the baseline SST model overestimates it. The $\beta$-Transport model displays a more diffused downstream $\beta$ distribution, resulting in a stronger correction compared to the SST-CND model. Figure 12 illustrates that both models forecast a more comprehensive velocity profile, with the $\beta$-Transport model aligning more closely with the LES data due to its extensive correction. Moreover, Figure 13 highlights that both models significantly decrease the reattachment point error compared to the SST model, indicating the effectiveness of the $\beta$-Transport model's data-driven calibration. The $\beta$-Transport model slightly underestimates the $x_{rth,HUMP}$, while the SST-CND model slightly overestimates it. However, in the recirculation zone and downstream, the $\beta$-Transport model predicts higher friction compared to the experiment, as its calibration is focused on reattachment rather than the entire $C_f$ distribution, and due to the absence of a non-linear Reynolds stress-mean flow relationship. In both Figure 12 and Figure 13, the results obtained using the three different levels of grids are quite similar, indicating the good grid convergence of the $\beta$-Transport model.

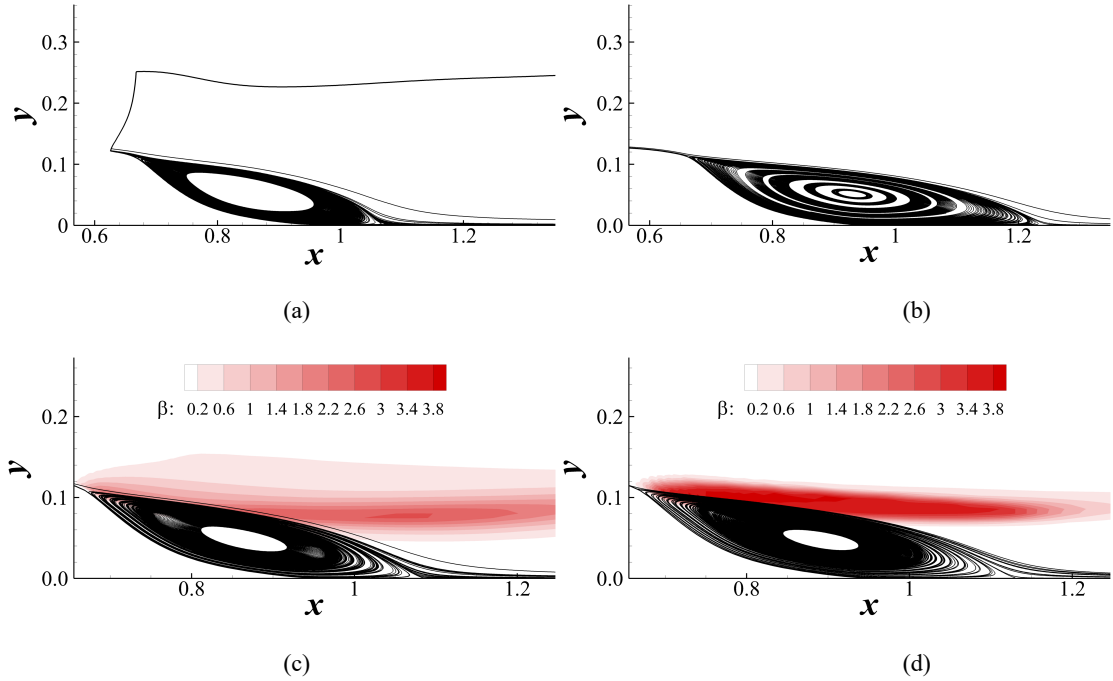

Figure 11. The streamline plots given by (a) LES[47], (b) the baseline SST model, (c) the $\beta$-Transport model, and (d) the SST-CND model; All the plots are obtained from the medium mesh.

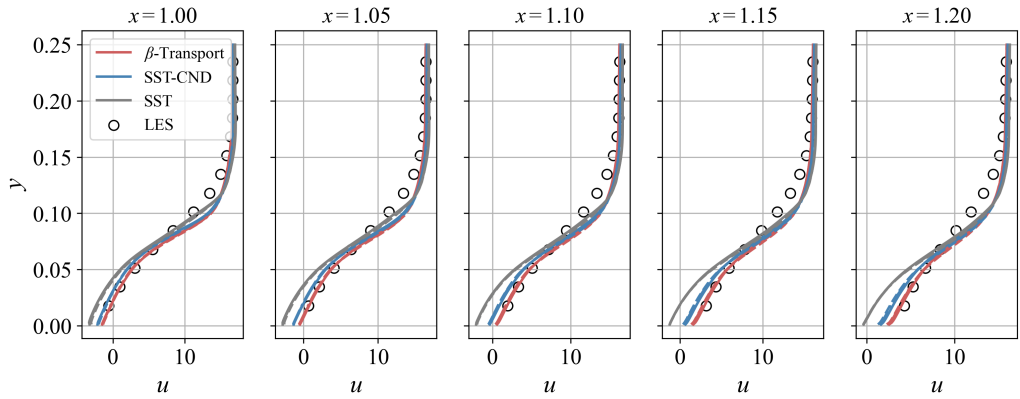

Figure 12. The velocity profiles given by different models. The dashed lines, dash-dot lines, and solid lines represent the result of the coarse mesh, medium mesh, and fine mesh, respectively.

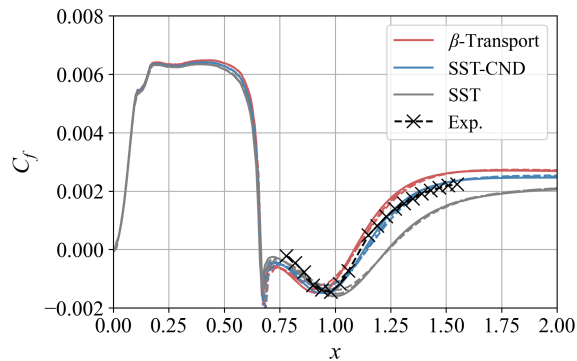

Figure 13. The friction coefficient predicted by different models. The dashed lines, dash-dot lines, and solid lines represent the result of the coarse mesh, medium mesh, and fine mesh.

In terms of velocity MSE, as shown in Table 5, the non-local $\beta$-Transport model demonstrates superior performance compared to the SST-CND model with local correction. This indicates that the $\beta$-Transport model more accurately captures the characteristics of the mean flow field.

Table 5. The velocity MSE / the velocity MSE of the SST model, fine mesh

|  | SST | SST-CND | $\beta$-Transport |
| --- | --- | --- | --- |
| MSE/MSE$_{SST}$ | 100% | 41.4% | **37.1%** |

### 3.4.3. Zero pressure gradient flat plate

The $\beta$-Transport model was also implemented in the optimization process for the zero-pressure gradient flat plate to evaluate its effectiveness. In this specific scenario, the SST model already demonstrates high accuracy. Therefore, it is necessary to ensure that the updated model does not compromise its performance. Figure 14(a) illustrates that the $C_f$ distributions predicted by the non-local $\beta$-Transport model, the local SST-CND model, and the SST model all align closely with the experimental data. Additionally, Figure 14(b) demonstrates that the velocity profiles generated by the $\beta$-Transport model closely resemble those of the SST model, displaying the typical logarithmic correlation with $y^+$ within the log layer. Consequently, the non-local adjustment in the $\beta$-Transport model does not negatively affect the accuracy of the baseline SST model's accuracy in wall-attached flows, indicating strong reliability.

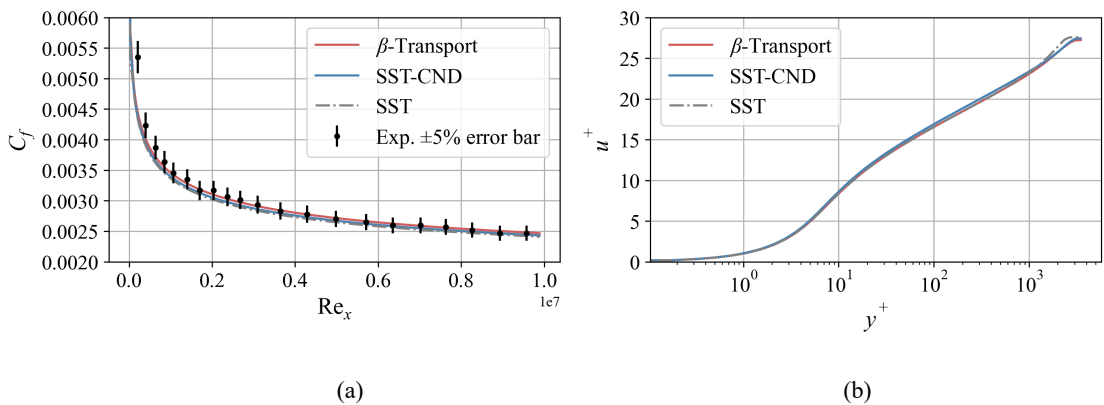

(a)  (b)

Figure 14. (a) $C_f$ distribution along the flat plate, (b) the velocity profile.

# 4. Test cases

In this section, we demonstrate the performance of the proposed non-local $\beta$-Transport model in test cases that are distinct from the training set. Our findings indicate that the $\beta$-Transport model yields a smaller velocity MSE compared to the local SST-CND model, highlighting its superior generalizability. Unless stated otherwise, all test cases were simulated using OpenFOAM's SimpleFoam solver.

## 4.1. Curved backward-facing step

The geometry and mesh [53][54] of the curved backward-facing step (CBFS) are shown in Figure 15. A total of 37,093 cells were utilized. In Figure 16), it can be observed that the baseline SST model significantly overestimates the size of the separation bubble when compared to the LES data, which serves as the reference. The local correction implemented in the SST-CND model does lead to some improvements by predicting a smaller separation size (as shown in Figure 16(b)), however, there remains a discrepancy with the LES data. Conversely, the proposed non-local $\beta$-Transport model accurately predicts a separation size that closely matches the LES data, as illustrated in Figure 16(c). By comparing Figure 16(c) with Figure 16(d), it is evident that the $\beta$ predicted by the proposed non-local model extends further downstream, providing a more pronounced correction compared to the SST-CND model. The velocity profiles presented in Figure 17 further validate the superior performance of the non-local $\beta$-Transport model over the local SST-CND correction and the baseline SST model, showcasing a more comprehensive velocity distribution. Additionally, Figure 18 demonstrates that the $\beta$-Transport model predicts a reattachment point that aligns closest with the LES data.

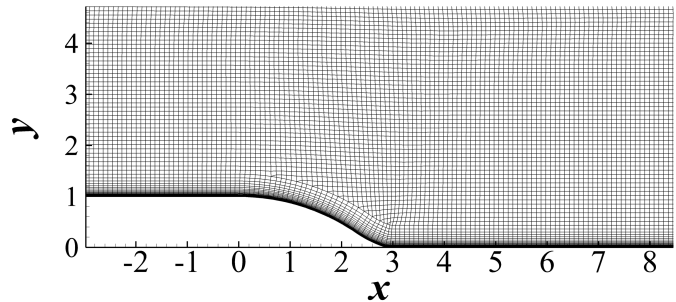

Figure 15. The geometry and computational mesh for the CBFS case.

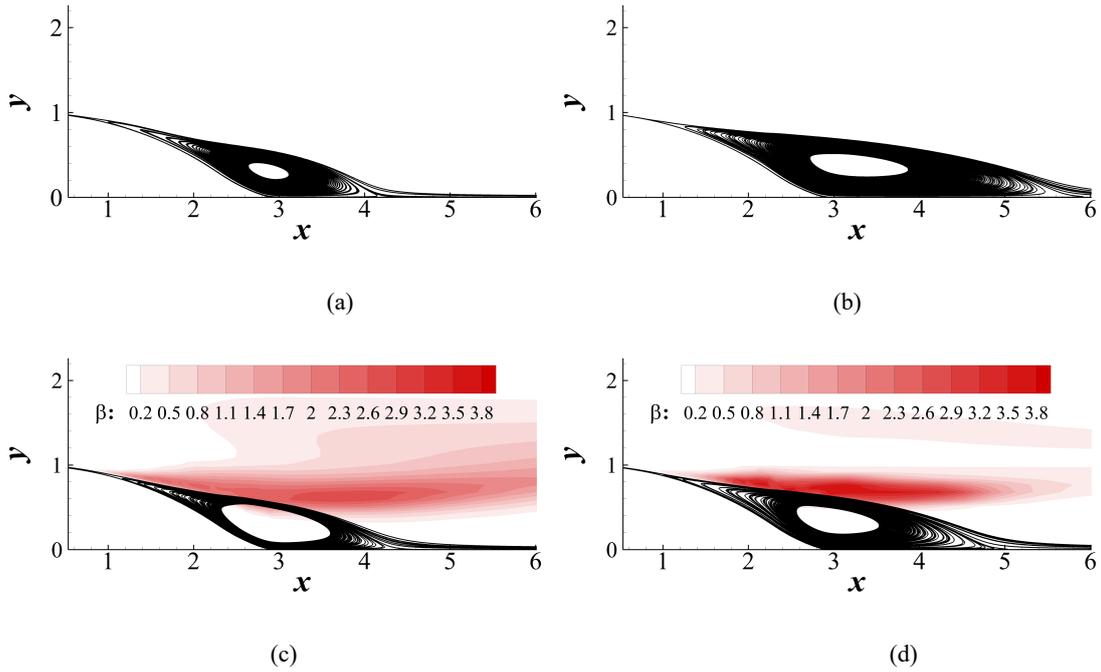

Figure 16. The streamline plot of the CBFS case given by (a) the LES data [53][54], (b) SST model, (c) $\beta$-Transport model, and (d) SST-CND model.

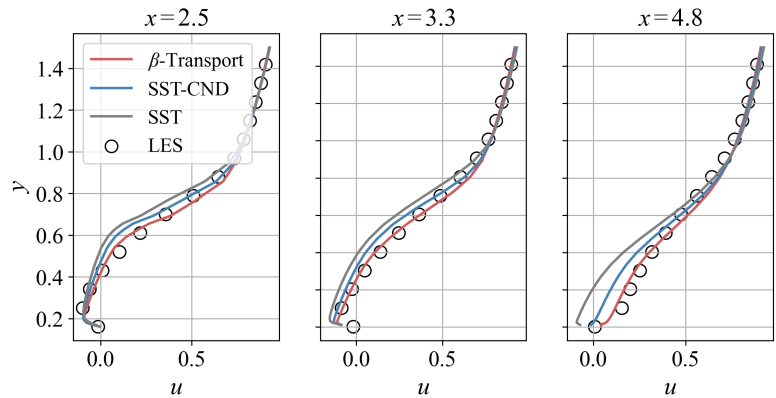

Figure 17. The velocity profiles at different locations.

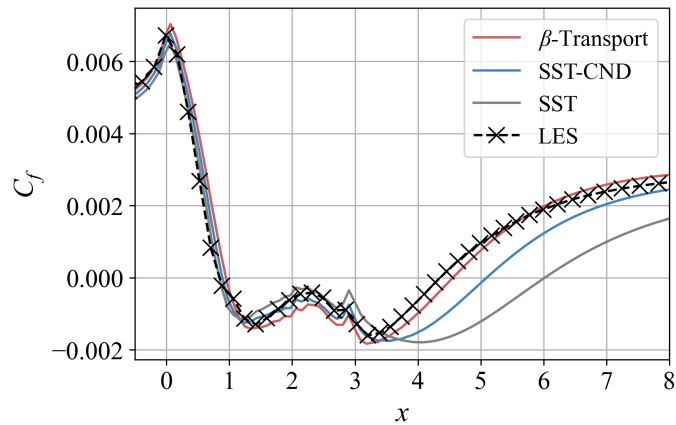

Figure 18. The $C_f$ distribution along the separation zone.

The velocity MSE in the separation zone was calculated using the Large Eddy Simulation (LES) data as a reference and is summarized in Table 6. The findings indicate that by including the non-local effect, the $\beta$-Transport model decreases the velocity MSE by approximately 20%, demonstrating greater overall accuracy compared to the SST-CND model, which only includes a local correction.

Table 6. The velocity MSE / the velocity MSE of the SST model in the CBFS case

|  | SST | SST-CND | $\beta$-Transport |
|---|---|---|---|
| MSE/MSE$_{SST}$ | 100% | 27.7% | **7.2%** |

## 4.2. Periodic hills with different aspect ratios and Reynolds numbers

The periodic hills with $\alpha = 0.8, 1.0, 1.2, 1.5$ from Xiao et al. [46] serve as the test cases in this section. The Reynolds number, based on the mean velocity at the inlet and the height of the hill, is 5600, which differs from the training case ($\text{Re}_h = 10595$). To ensure grid convergence, three levels of meshes are utilized for each $\alpha$ value. In the following paragraphs, we will initially present the results for the $\alpha = 0.8$ case, and subsequently touch on the other cases briefly. Figure 19 illustrates that the baseline SST model notably overestimates the separation size in the periodic hill with $\alpha = 0.8$. However, both the corrections in the $\beta$-Transport model and the SST-CND model exhibit enhancements over the baseline SST model. Notably, the non-local correction in the $\beta$-Transport model results in a smaller separation bubble, aligning better with the high-fidelity DNS data. The velocity profiles in Figure 20 further indicate that the $\beta$-Transport model produces results that closely match the LES data. The consistent profiles across different meshes highlight the favorable mesh convergence. Moreover, the $C_f$ distributions presented in Figure 21 underscore that the non-local correction within the $\beta$-Transport model significantly improves the accuracy of the predicted reattachment point. It is worth noting that the $C_f$ data for Xiao's periodic hills has not been disclosed. Therefore, we utilized the disclosed DNS mean velocity data with OpenFOAM's post-processing function to obtain high-fidelity data for reference.

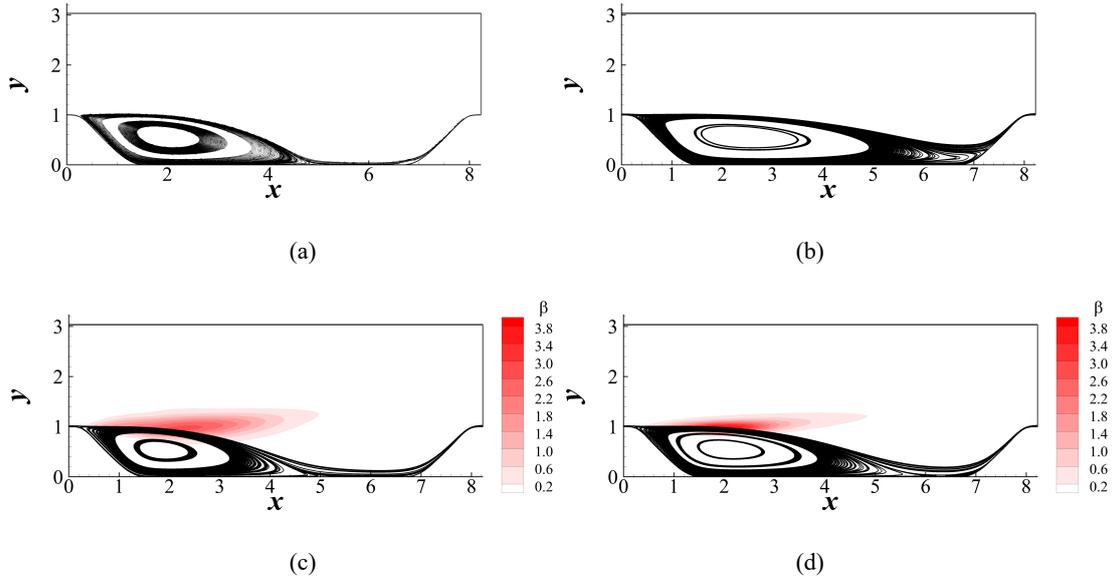

Figure 19. The streamline plots of the periodic hill with $\alpha = 0.8$ given by (a) the DNS data, (b) SST model, (c) $\beta$-Transport model, and (d) SST-CND model; the results plotted are from the medium mesh.

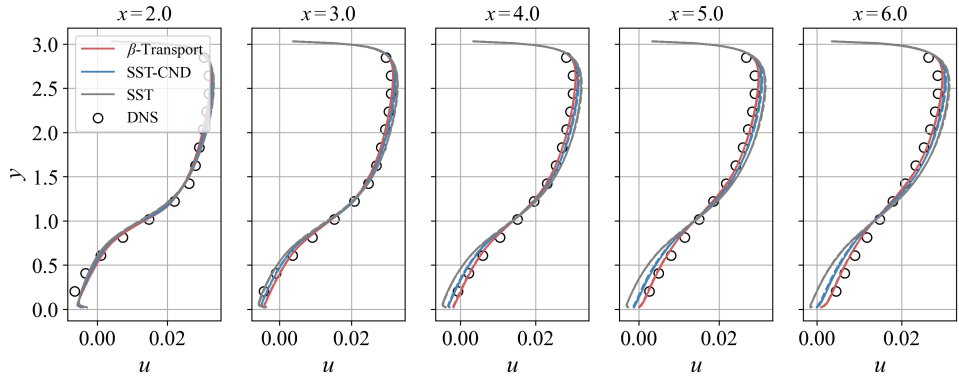

Figure 20. The velocity profiles given by different models; the periodic hill with $\alpha = 0.8$. The dashed lines, dash-dot lines, and solid lines represent the result of the coarse mesh, medium mesh, and fine mesh.

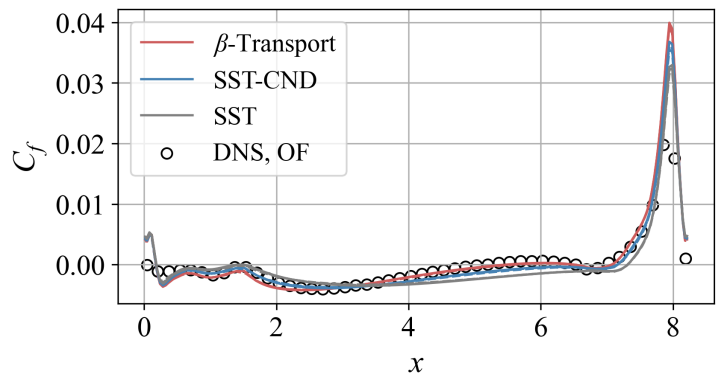

Figure 21. The $C_f$ distribution given by different models; the periodic hill with $\alpha = 0.8$. The dashed lines, dash-dot lines, and solid lines represent the result of the coarse mesh, medium mesh, and fine mesh.

The streamline plots and the $C_f$ distributions are very similar for the other cases with $\text{Re}_h = 5600$. Hence, they are omitted to keep the manuscript compact. The velocity profiles of these cases are shown in Figure 22(a) to Figure 22(c), all indicating that the $\beta$-Transport model with non-local correction performs the best. Grid convergence is also confirmed. The MSE listed in Table 7 demonstrates that the $\beta$-Transport model exhibits better generalizability in the periodic hill case compared to the SST-CND model.

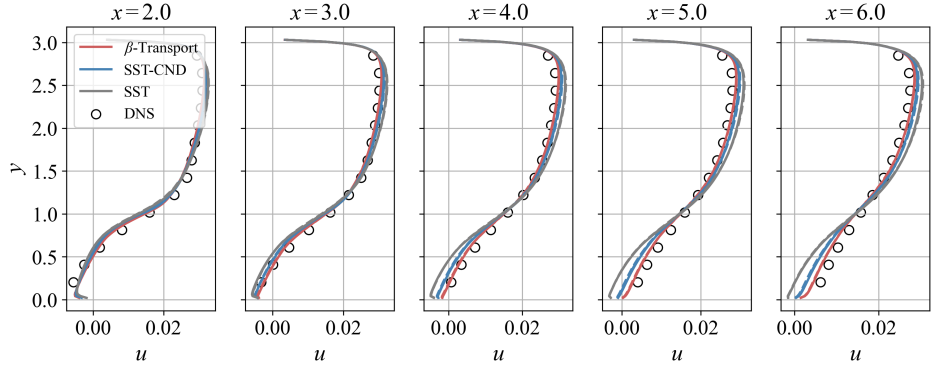

(a)

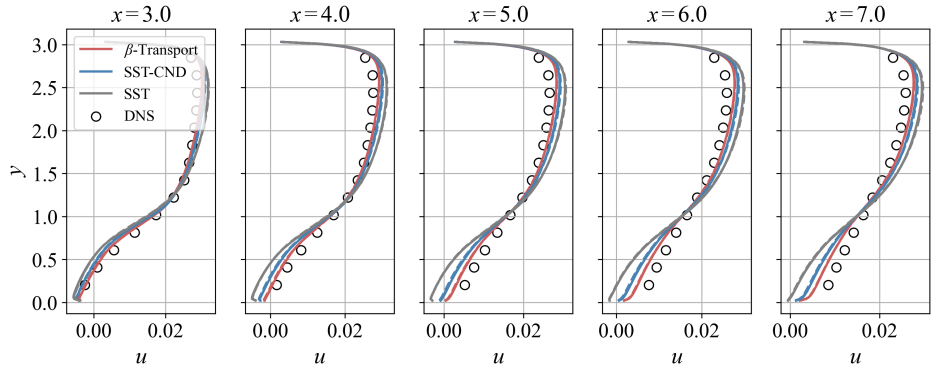

(b)

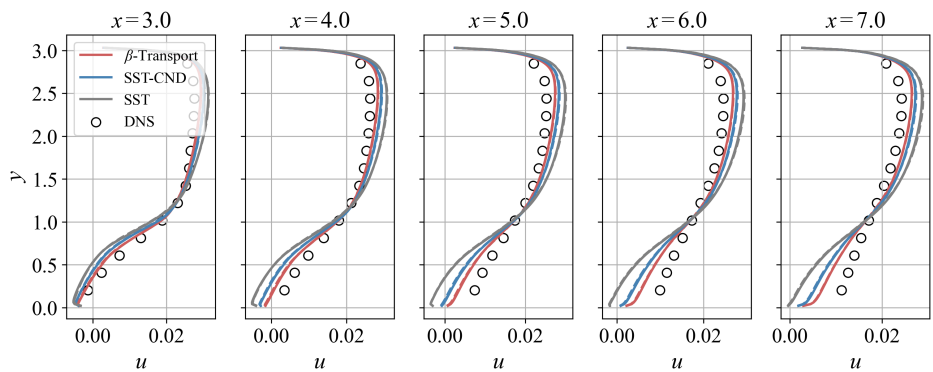

(c)

Figure 22. The velocity profiles given by different models, the periodic hill with $\text{Re}_h = 5600$, (a) $\alpha = 1.0$, (b)

$\alpha = 1.2$, (c) $\alpha = 1.5$; results from different mesh levels are represented using different line styles.

Table 7. The velocity MSE / the velocity MSE of the SST model in the periodic hill cases

| $MSE/MSE_{SST}$ | SST | SST-CND | $\beta$-Transport |
|---|---|---|---|
| $Re_h = 5600, \alpha = 0.8$ | 100% | 28.9% | **13.3%** |
| $Re_h = 5600, \alpha = 1.0$ | 100% | 29.7% | **8.8%** |
| $Re_h = 5600, \alpha = 1.2$ | 100% | 32.9% | **11.6%** |
| $Re_h = 5600, \alpha = 1.5$ | 100% | 39.8% | **17.4%** |

## 4.3. 2D-bump case

In this section, the non-local $\beta$-Transport model is applied to the 2D-bump case with $h = 31\ mm$. The geometry and computational mesh (coarse) are shown in Figure 23 [53][55]. The case exhibits a mild flow separation downstream of the bump, as shown in Figure 24(a). This separation is not as pronounced as the extensive separation observed in the $\beta$-Transport model training data, making this case particularly challenging to compute. The coarse mesh, medium mesh, and the fine mesh have $293 \times 125$, $412 \times 175$, and $581 \times 249$ cells, respectively. Figure 24(b) illustrates that the baseline SST model significantly underestimates the size of the separation zone. Comparing Figure 24(c) and (d), it can be observed that both the non-local $\beta$-Transport model and the local SST-CND model reduce the separation size compared to the baseline SST model. Despite still showing delayed reattachment and premature separation, the $\beta$-Transport model exhibits a milder separation compared to the SST-CND model, bringing it closer to matching the LES data. Similarly to previous test cases and training cases, the non-local correction implemented in the transport equation of the $\beta$-Transport model provides a more diffuse and extensive correction than the local correction used in the SST-CND model. Velocity profiles displayed in Figure 25 confirm that the results obtained by the $\beta$-Transport model are the closest to the high-fidelity LES data. The overlapping profiles from different meshes indicate a good grid convergence property. The velocity MSE (with LES data as the reference) is presented in Table 8. Notably, the MSE drops considerably when using the $\beta$-Transport model compared to the SST-CND model, highlighting the enhancements introduced by the non-local effect.

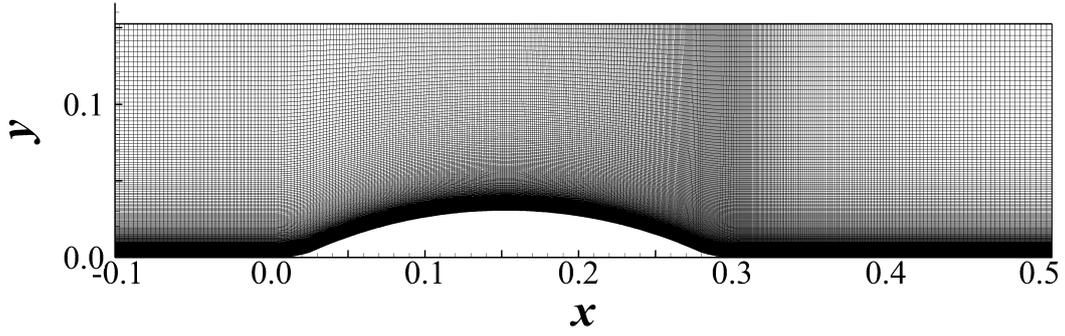

Figure 23. The domain and the computational mesh of the 2D-bump case.

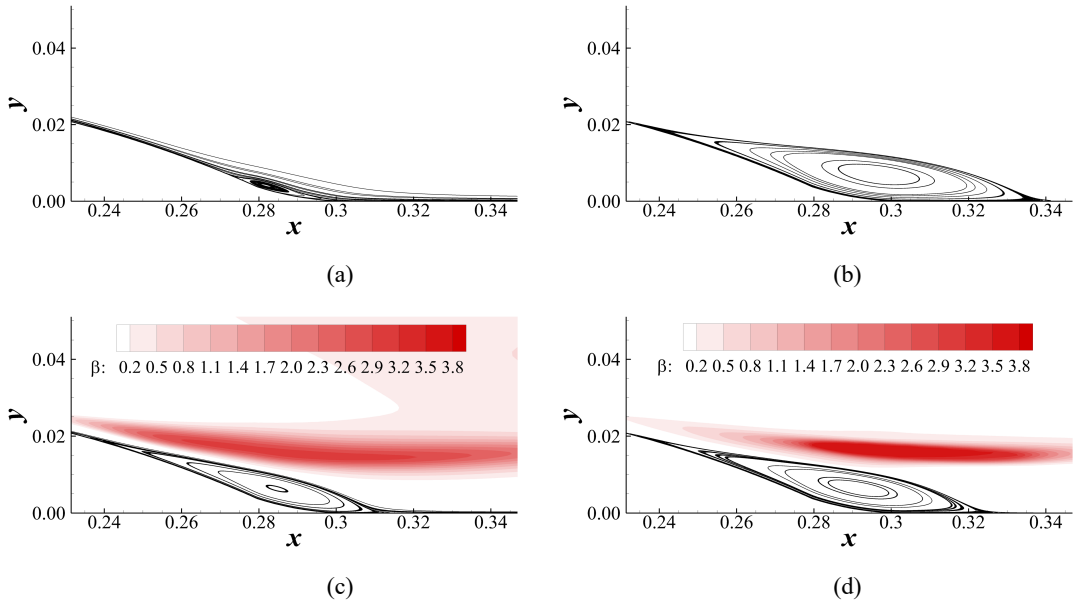

Figure 24. The streamline plot given by (a) the LES data, (b) baseline SST model, (c) $\beta$-Transport model, and (d) the SST-CND model; the results plotted are from the medium mesh.

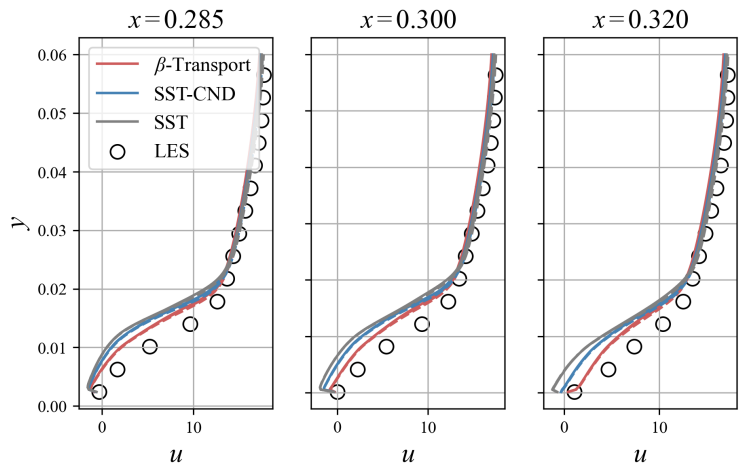

Figure 25. The velocity profiles given by different models. The dashed lines, dash-dot lines, and solid lines represent the coarse mesh, medium mesh, and fine mesh, respectively.

Table 8. The velocity MSE / the velocity MSE of the SST model in the 2D-bump case

|  | SST | SST-CND | $\beta$-Transport |
|---|---|---|---|
| MSE/MSE$_{SST}$ | 100% | 64.4% | **36.5%** |

## 4.4. Ahmed body

In this section, the three-dimensional Ahmed body, as proposed in [56], is examined using the $\beta$-Transport model. Figure 26 illustrates the configuration and computational domain. It is important to note that only half of the configuration is analyzed. The structured mesh, depicted in Figure 27, is utilized for this study. To ensure grid convergence, both a coarse mesh consisting of 3.6 million cells and a fine mesh with 7 million cells are employed. The freestream velocity is set at $40\ m/s$ and the Reynolds number based on the length of the body is $\text{Re}_L = 2.78 \times 10^6$. Figure 28 shows the streamline and contour of $\beta$ on the symmetry plane. The results suggest that the SST model indicates a separation beginning at the slant, while both the SST-CND and the $\beta$-Transport model exhibit separation at the rear of the body. Notably, the streamlines produced by the SST-CND and $\beta$-Transport models are quite similar. Furthermore, Figure 28(b) and (c) demonstrate that the $\beta$ distribution generated by the transport equations-based model is more diffuse and smoother. In Figure 29, the velocity profiles are compared to the experimental results conducted previously [57] on the symmetry plane. The findings support that both the SST-CND model and $\beta$-Transport model align more closely with the experimental data. Additionally, there is good agreement between the velocity profiles obtained from the coarse and fine meshes, suggesting favorable grid convergence properties. The iso-surface of the Q-criterion is depicted in Figure 30, showing that the three-dimensional flow structures predicted by the SST-CND model and $\beta$-Transport model exhibit similarities. The SST model displays a significant 3-D separation originating from the slant. Moreover, the distribution of turbulent kinetic energy $k$ on the iso-surfaces indicates that the corrections applied by the SST-CND model and $\beta$-Transport model tend to enhance turbulent intensity in the separated shear layers.

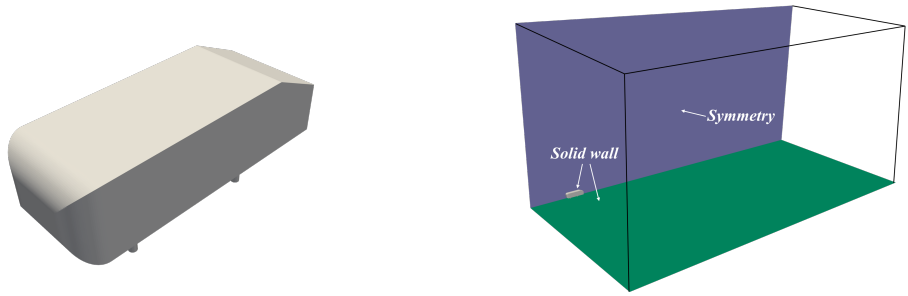

Figure 26. (a) The configuration and (b) the computational domain; all other far-field patches are set to an inlet-outlet boundary condition.

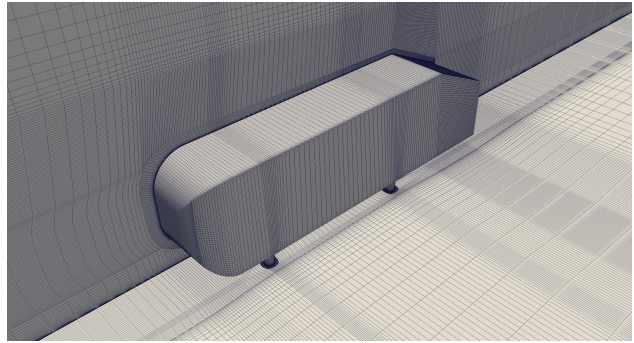

Figure 27. The computational mesh (coarse).

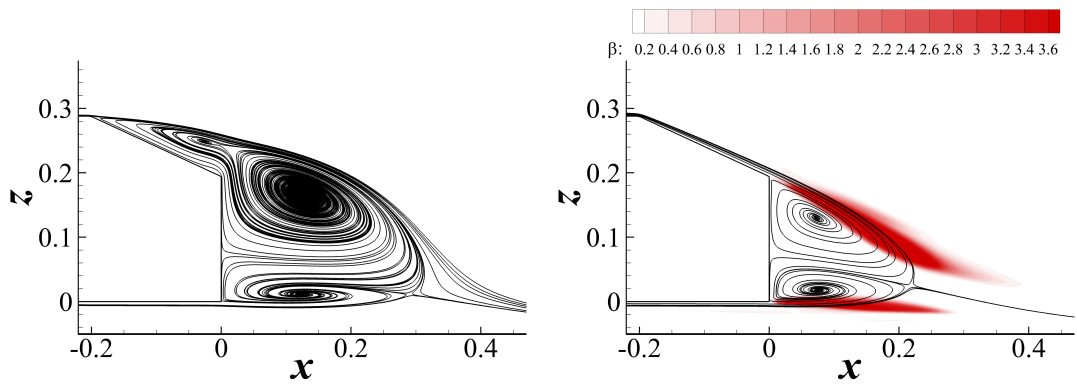

(a)          (b)

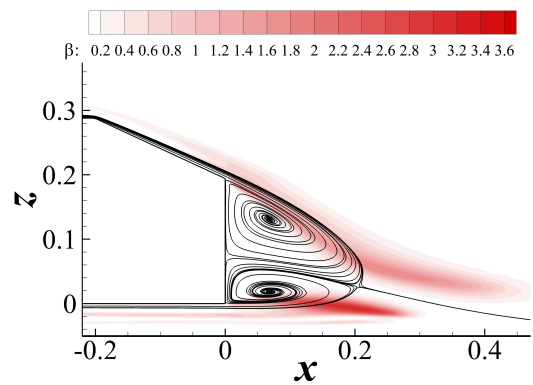

(c)

Figure 28. The streamline plot given by (a) the SST model, (b) SST-CND model, (c) $\beta$-Transport model.

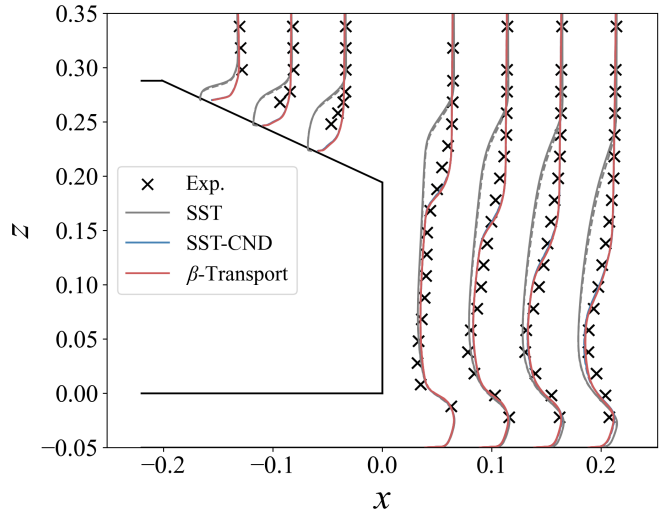

Figure 29. The velocity profiles given by different models. The results of the coarse mesh are represented by dashed lines, while the results from the fine mesh are represented by solid lines.

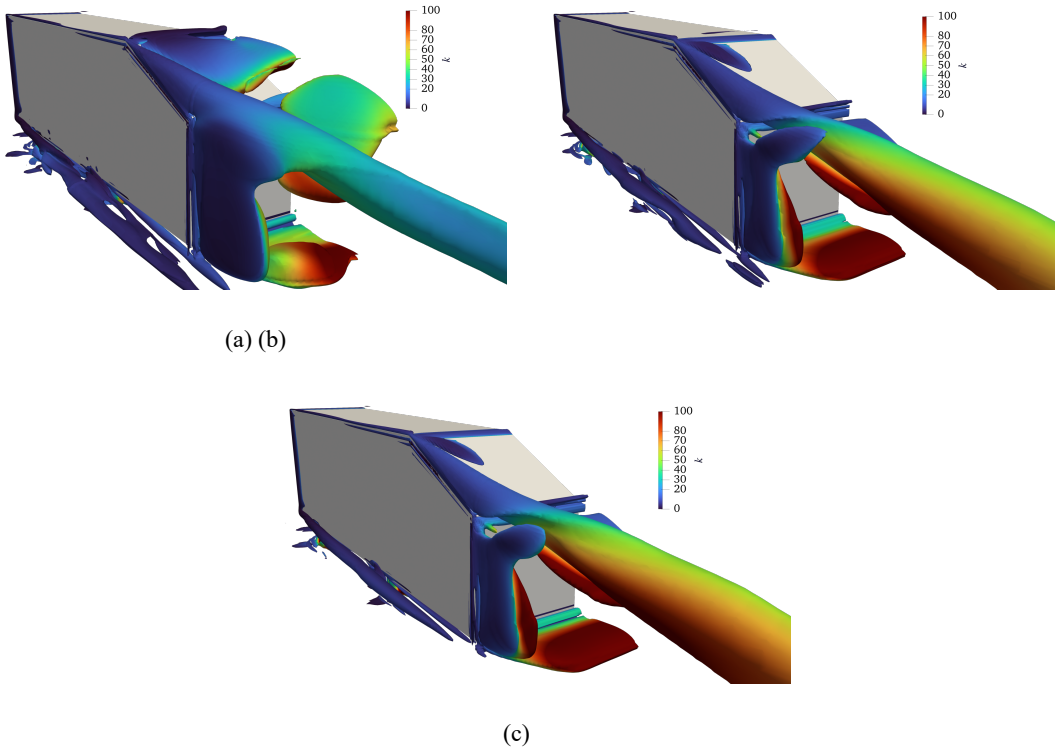

(a)  (b)

(c)

Figure 30. The iso-surface of the Q-criterion determined by (a) the SST model, (b) SST-CND model, (c) $\beta$-Transport model; the iso-surfaces are colored based on the turbulent kinetic energy $k$.

In conclusion, both the SST-CND model and the $\beta$-Transport model demonstrate decent accuracy in predicting 3-D separated flows. The findings further suggest that the non-local transport equation-based model ($\beta$-Transport) introduced in this study is well-suited for 3-D calculations.

## 4.5. Basic flows

In this section, we analyze the performance of the $\beta$-Transport model in both mixing layer and channel flows. The computational mesh used for the 2-dimensional mixing layer flow is displayed in Figure 31. The mesh contains $1.7 \times 10^5$ cells and is adopted from NASA's second-finest mesh for the mixing layer test case [58]. The lengths of the upper and lower plates have been adjusted to align with the velocity profile 1 mm downstream of the splitter plate's tail, matching the experimental data. The freestream velocities are set to 41.54 m/s for the upper plate and 22.40 m/s for the lower plate. The Reynolds number, based on the upper plate's freestream velocity and a reference length of $L = 1\ mm$ is approximately 2900. The streamwise coordinate $x$ is defined as zero at the end of the splitter plate. Figure 32 presents the velocity profiles at $x = 1$ mm, indicating a close match between the computed velocity distribution at the start of mixing and the experimental data. The growth of vorticity thickness is illustrated in Figure 33(a), where the vorticity thickness is defined as:

$$\delta_\omega = \frac{U_a - U_b}{\left(\frac{\partial u}{\partial y}\right)_{max}}, U_a = 41.54\ \text{m/s}, U_b = 22.40\ \text{m/s} \quad (19)$$

$U_a$ and $U_b$ are the free stream velocities of the upper plate and the lower plate, respectively. Figure 33(a) indicates that the growth rate from both the $\beta$-Transport model and the SST model align well with the experimental data, whereas the SST-CND model shows a significantly larger growth rate, which is erroneous. However, the $\beta$-Transport model still produces a greater absolute thickness compared to the experimental data. Moving on to Figure 33(b), the normalized velocity profiles generated by all three models closely match the experimental data, indicating the achievement of a self-similar state. Figure 34 reveals that the $\beta$-Transport model predicts a very small $\beta$ across most of the mixing region, while the SST-CND model consistently yields $\beta > 0$ throughout the entire mixing region. This discrepancy accounts for the deviated growth rate of the vorticity thickness predicted by the SST-CND model.

The discrepancy between the $\beta$-Transport model and the SST-CND model in the mixing layer case can be elucidated at the equation level. In the $\beta$-Transport model, as per Eq. (8), the production term is inversely proportional to $1/d^2$, leading it to approach 0 as the distance from the splitter plate increases within the mixing region. This limitation restrains the growth of $\beta$ downstream of the splitter plate (even though Eq. (8) does not guarantee that $\beta$ strictly equals zero downstream), resulting in minimal impact on the original SST model. Conversely, in the SST-CND model, the correction term's magnitude is directly linked to the local shear (rotation), resulting in an increased $\beta$ (from 0) across the entire mixing region since shear is prevalent in this area.

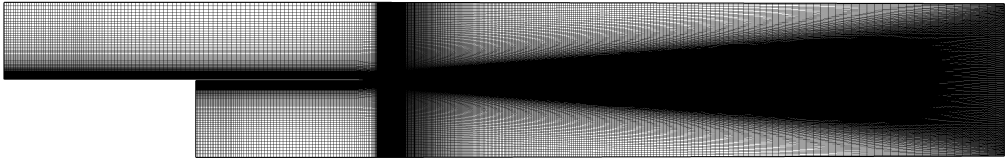

Figure 31. The computational domain and mesh of the mixing layer.

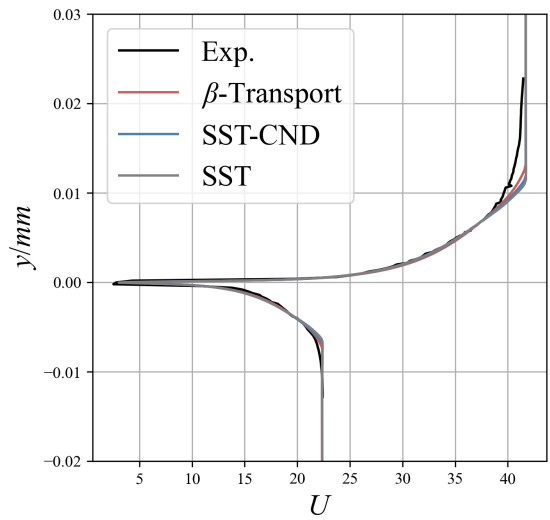

Figure 32. The velocity profiles at $x = 1$ mm.

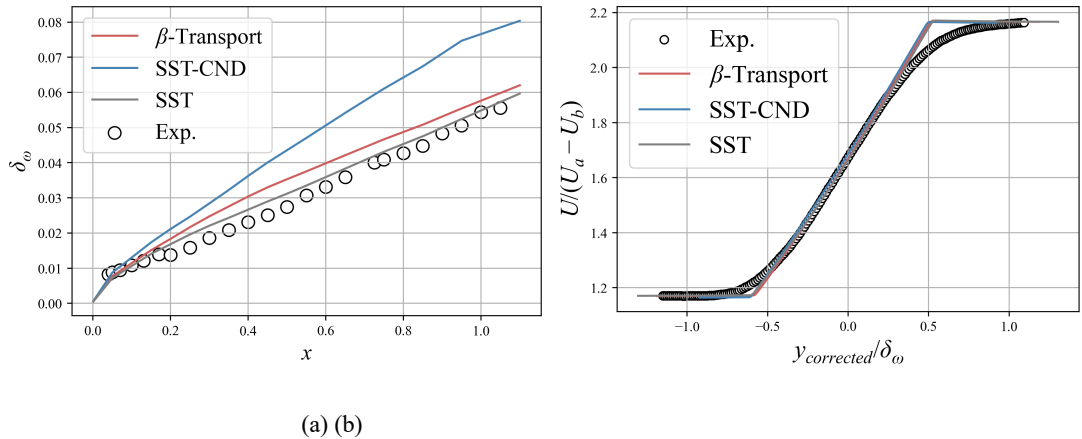

(a) (b)

Figure 33. (a) The growth of the vorticity thickness along the streamwise direction; (b) the self-similar velocity profile, where $U_a = 41.54 \ m/s$, $U_b = 22.4 \ m/s$ and $\delta_\omega$ is the vorticity thickness.

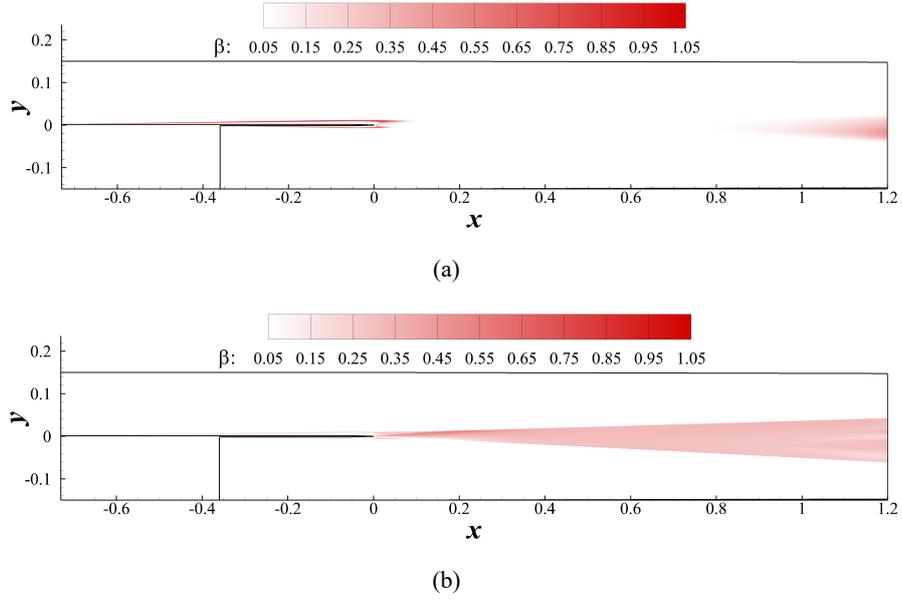

Figure 34. The $\beta$ contour given by (a) the $\beta$-Transport model and (b) the SST-CND model.

The channel flows with $\text{Re} = 6.7 \times 10^4, 6.7 \times 10^5, 1.2 \times 10^6$ (based on the mean velocity and the half-channel height) are also investigated using the new $\beta$-Transport model. OpenFOAM's 1-dimensional solver BoundaryFoam is utilized for this study. The height of the first layer ensures $\Delta y^+ < 0.01$. The profiles of the normalized quantities (using the friction velocity $u_\tau$ and the inner length scale $\delta_v$) are plotted in Figure 35. The results in Figure 35 show that the profiles predicted by the $\beta$-Transport model, the SST-CND model, and the SST model are nearly identical for the same Reynolds number. Therefore, the new $\beta$-Transport model does not negatively affect the accuracy of the baseline SST model for basic channel flows. The same conclusion applies to the SST-CND model.

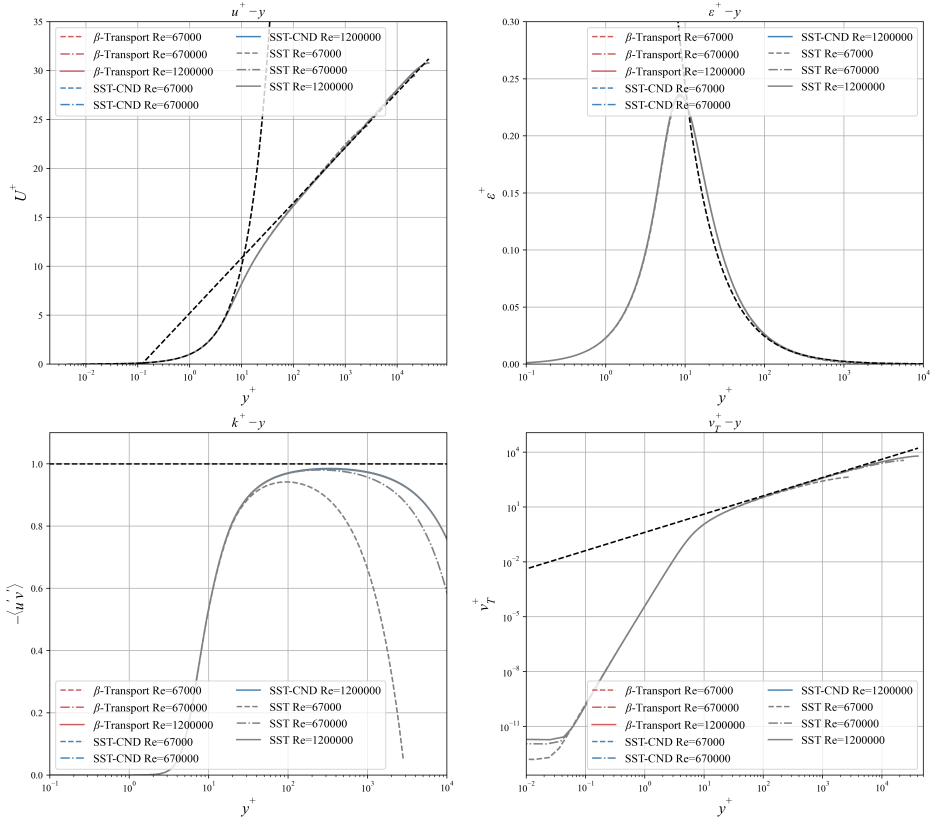

Figure 35. The profiles of $u^+, \varepsilon^+, k^+,$ and $\nu_T^+$ given by different model; the profiles for the same Reynolds number exhibit significant overlap.

In conclusion, the $\beta$-Transport model has been shown to maintain the accuracy of the baseline SST model in both simple wall-attached flows and free shear flows, such as the mixing layer. Specifically, the non-local formation of the $\beta$-Transport model results in a more realistic spreading rate of the mixing layer when compared to the SST-CND model, which relies solely on local shear.

## 5. Conclusions

In this study, non-local characteristics are partially integrated into the data-driven RANS turbulence modeling framework by introducing a transport equation for the correction term. The resulting $\beta$-Transport model is calibrated using the differential evolutionary algorithm with a response surface based on the high-fidelity data of NASA hump and the periodic hill with $\alpha = 0.8$. Subsequent tests on separated flows demonstrate that the $\beta$-Transport model (using a non-local correction) outperforms the SST-CND model in [33] (using a local correction) in terms of velocity MSE. Moreover, in fundamental flows such as the mixing layer and the channel flow, the $\beta$-

Transport model demonstrates similar accuracy compared to the baseline SST model. Our findings suggest that considering the non-local nature of turbulence in the data-driven RANS turbulence modeling framework can enhance the model's generalizability in complex separated flows, providing a new avenue for training models with superior performance.

In this study, the transport equation for the correction term has been developed empirically, with only the model parameters calibrated through data-driven methods. Further research can explore deriving transport equations directly from data using more advanced techniques.

# Acknowledgement

This work was supported by the National Natural Science Foundation of China (grant nos. 12372288, 12388101, U23A2069, and 92152301). The authors are sincerely grateful for the constructive discussions with Prof. Xiang I.A. Yang of Pennsylvania State University.

# Declaration of interests

The authors report no conflict of interest.